\documentclass[showpacs,amsmath,amssymb,twocolumn,superscriptaddress,notitlepage,preprintnumbers,pra,floatfix,hyperref]{revtex4-1}

\usepackage{qcircuit} 
\usepackage[dvips]{graphicx}
\usepackage{amsmath,amssymb,amsthm,mathrsfs,amsfonts,dsfont}
\usepackage{subfigure, epsfig}
\usepackage{braket}
\usepackage{bm}
\usepackage{enumerate}
\usepackage{color}
\usepackage{graphicx}
\usepackage{algorithm}
\usepackage{algpseudocode}
\usepackage{braket}
\usepackage{xr}
\usepackage{hyperref}
\hypersetup{colorlinks=true, linkcolor=blue, citecolor=blue, urlcolor=black }

\newcommand{\tr}{\mathrm{Tr}}

\newcommand{\mc}[1]{\mathcal{#1}}

\newcommand{\sun}[1]{\textcolor{black}{#1}}
\newcommand{\hl}[1]{\textcolor{black}{#1}}
\newcommand{\zt}[1]{\textcolor{black}{#1}}

\newcommand{\vabs}[1]{\left\| #1 \right\|}
\newcommand{\pbra}[1]{\left( #1 \right)}
\newcommand{\cbra}[1]{\left\{ #1 \right\}}
\newcommand{\sbra}[1]{\left[ #1 \right]}

\newcommand{\Kcal}{\mathcal{K}}

\newcommand{\bmat}{\bm{\mathrm b}}
\newcommand{\Omat}{\bm{\mathrm O}}
\newcommand{\Pmat}{\bm{\mathrm P}}
\newcommand{\Qmat}{\bm{\mathrm Q}}
\newcommand{\Olmat}{\bm{\mathrm O}_l}
\newcommand{\Olpmat}{\bm{\mathrm O}_{l'}}

\newcommand{\Plmat}{\bm{\mathrm P}_l}

\newcommand{\abs}[1]{\left| #1 \right|}

\newcommand{\QST}[1]{\langle #1 \rangle^{\text{QST}}}
\newcommand{\ES}[1]{\langle #1 \rangle^{\text{ES}}}

\newcommand{\RNum}[1]{\uppercase\expandafter{\romannumeral #1\relax}}

\newcommand{\omat}{\bm{\mathrm o}}

\begin{document}

\title{Experimental quantum state measurement with classical shadows}

\date{\today}

\author{Ting Zhang}
\thanks{These two authors contributed equally }
\affiliation{School of Physics, Shandong University, Jinan 250100, China}

\author{Jinzhao Sun}
\thanks{These two authors contributed equally }
\affiliation{Center on Frontiers of Computing Studies,  Peking University, Beijing 100871, China}
\affiliation{Clarendon Laboratory, University of Oxford, Parks Road, Oxford OX1 3PU, United Kingdom}

\author{Xiao-Xu Fang}
\affiliation{School of Physics, Shandong University, Jinan 250100, China}

\author{Xiao-Ming Zhang}
\affiliation{Center on Frontiers of Computing Studies,  Peking University, Beijing 100871, China}
\affiliation{Department of Physics, City University of Hong Kong, Tat Chee Avenue, Kowloon, Hong Kong SAR, China}

\author{Xiao Yuan}
\email{xiaoyuan@pku.edu.cn}
\affiliation{Center on Frontiers of Computing Studies,  Peking University, Beijing 100871, China}

\author{He Lu}
\email{luhe@sdu.edu.cn}
\affiliation{School of Physics, Shandong University, Jinan 250100, China}


\begin{abstract}
A crucial subroutine for various quantum computing and communication algorithms is to efficiently extract different classical properties of quantum states. In a notable recent theoretical work by Huang, Kueng, and Preskill~[Nat.~Phys.~16,~1050~(2020)], a thrifty scheme showed how to project the quantum state into classical shadows and simultaneously predict $M$ different functions of a state with only $\mathcal {O}(\log_2 M)$ measurements, independent of the system size and saturating the information-theoretical limit. Here, we experimentally explore the feasibility of the scheme in the realistic scenario with a finite number of measurements and noisy operations. We prepare a four-qubit GHZ state and show how to estimate expectation values of multiple observables and  Hamiltonians. We compare the measurement strategies with uniform, biased, and derandomized classical shadows to conventional ones that sequentially measure each state function exploiting either importance sampling or observable grouping.
We next demonstrate the estimation of nonlinear functions using classical shadows and analyze the entanglement of the prepared quantum state.
Our experiment verifies the efficacy of exploiting (derandomized) classical shadows and sheds light on efficient quantum computing with noisy intermediate-scale quantum hardware. 
\end{abstract}
 
\maketitle

 \emph{{Introduction.}---} 
Quantum computers could process information in parallel and efficiently represent many-body quantum states~\cite{mcardle2018quantum,cerezo2020variational,endo2020hybrid,PhysRevX.8.011021}. Yet, the power of quantum computing subjects us to how efficiently we extract classical information from the quantum state.
Focusing on variational quantum algorithms designed for near-term quantum devices~\cite{peruzzo2014variational,PhysRevX.8.011021,cerezo2020variational,bharti2021noisy,mcardle2018quantum,huggins2021efficient,endo2020variational,endo2020hybrid,hempel2018quantum,yuan2019theory,google2020hartree,yuan2020quantum,fujii2020deep,commeau2020variational,bravo-prieto2019,XU2021,huang2019near,mcardle2018quantum}, whether they are sufficiently effective to demonstrate clear and robust quantum advantages, relies on how efficiently we can measure the state~\cite{mcclean2016theory,Cao2019Quantum,izmaylov2019revising,rubin2018application,o2019generalized,Yen2019,izmaylov2019unitary,zhao2020measurement,huggins2021efficient,sun2021perturbative,Yen2019,Gokhale2019,torlai2018neural,carrasquilla2019reconstructing,torlai2020precise}. For example, the Hamiltonian of a molecule with $M$ modes has  $\mc O(M^4)$ terms and a naive strategy requires $\mc O(M^8/\varepsilon^2)$ samples to measure each term to an accuracy $\varepsilon$~\cite{babbush2018low,mcardle2018quantum,google2020hartree}. In order to demonstrate a quantum advantage, we need to consider a sufficiently large $M$, say 100, and the cost of naively measuring those quantum systems could already be impractically large.

Advanced measurement schemes have been proposed to more efficiently evaluate observable expectation values without increasing the circuit depth~\cite{kandala2017hardware,verteletskyi2020measurement,huang2020predicting,wecker2015progress,hadfield2020measurements,huang2021efficient,torlai2018neural,choo2020fermionic,torlai2020precise,huang2020predicting,huang2021efficient,hadfield2020measurements,bonet2020nearly,cotler2020quantum,carrasquilla2019reconstructing}. One can use the strategy of importance sampling to  economically distribute more measurements to  observables with large contributions~\cite{wecker2015progress,mcclean2016theory}, or group compatible  observable  to reduce the cost in estimating low-weight qubit reduced density matrices~\cite{bonet2020nearly,cotler2020quantum} or observable expectations~\cite{kandala2017hardware,verteletskyi2020measurement,izmaylov2019unitary,zhao2020measurement,hempel2018quantum,o2016scalable}. 

Another notable scheme~\cite{huang2020predicting,aaronson2019shadow,struchalin2021experimental,chen2020robust,zhao2020fermionic,acharya2021informationally,hadfield2021adaptive,hillmich2021decision,garcia2021quantum} shows how to simultaneously obtain expectation values of multiple observables by randomly measuring and projecting the quantum state into classical shadows (CS). The algorithm only requires $\mc O(\log_2M)$ samples to measure $M$ low-weight observables, and the recently proposed locally biased CS~\cite{hadfield2020measurements} and derandomized CS~\cite{huang2021efficient} can be further applied to general observables with numerical results showing advantages over most other existing methods. 

While the advanced measurement schemes have been extensively studied in theory, their feasibility and comparison with realistic hardware are under exploration. In particular, efficiently implementing random measurements and analyzing how the noise in realistic hardware affects the measurement efficiency are critical for studying their practical performance with realistic devices. 

Here, we experimentally investigate the feasibility of the advanced measurement schemes with a four-qubit photonic quantum processor. We consider the schemes using importance sampling~\cite{wecker2015progress,mcclean2016theory}, observable grouping~\cite{kandala2017hardware,verteletskyi2020measurement,o2016scalable}, and the three schemes with uniformly random~\cite{huang2020predicting}, biased random~\cite{hadfield2020measurements}, and derandomized~\cite{huang2021efficient} classical shadows in tasks of estimating multiple local observables and computing the expectation of  Hamiltonian and its powers. We further apply the classical shadows to estimate the state purity and moments of the partially transposed density matrix, which helps to analyze its entanglement structure~\cite{He2018EntanglementStructure}. Our experimental results clearly show advantages of using (derandomized) classical shadows with realistic quantum devices. 

\emph{{Framework.}---}We first review the advanced measurement schemes in a unified framework recently proposed in Ref.~\cite{wu2021overlapped}. We aim to estimate the expectation value of an observable $\Omat$, which is decomposed into the Pauli basis as $\Omat = \sum_l \alpha_l \Olmat$ with $\Olmat \in \{I,X,Y,Z\}^{\otimes n}$ being the tensor product of single-qubit Pauli operators. For a multi-qubit Pauli operator $\Qmat = \otimes_{i=1}^n Q_i$ with $Q_i \in \{I, X,Y,Z\}$ being a single-qubit Pauli operator acting on the $i$th qubit, its expectation value can be obtained by measurements in any Pauli basis $\Pmat=\otimes_{i=1}^n P_i$ whenever $Q_i = P_i$ or $Q_i = I$ for any $i$, which we refer as $\Pmat$ hits $\Qmat$ and denote by $\Qmat\triangleright\Pmat$. When two Pauli observables are hit by the same basis $\Pmat$, we say that they are compatible with each other, and their expectation values can be simultaneously obtained by measuring the basis $\Pmat$. Considering two extreme cases of measuring $\Omat = \sum_l \alpha_l \Olmat$, the first one is that all the expectation values of $\Omat_l$ can be determined by one measurement $\Pmat$ if $\Olmat \triangleright \Pmat$ ($\forall l$)  i.e., every $\Omat_l$ is compatible with each other. On the contrary, we have to measure every $\Omat_l$ if no observable is compatible with any other one.

In general, to estimate $\tr(\rho\Omat)$ for an $n$-qubit unknown quantum state $\rho$, the measurement $\Pmat$ is randomly selected over the distribution $\Kcal(\Pmat)$.
An estimator for the target observable $\Omat$ is expressed as
\begin{equation}
\hat{\omat}(\Pmat) = \sum_l \alpha_l  f(\Pmat,\Olmat,\Kcal )\mu(\Pmat, \textrm{supp}(\Olmat))
\label{eq:estimator_main}
\end{equation}
where $\mu(\Pmat,\textrm{supp}(\Olmat)) = \prod_{i\in \textrm{supp}(\Olmat)}\mu(P_i)$ with $\mu(P_i)$ being the single-shot outcome of measurement $P_i$ on the $i$th qubit, $\textrm{supp}(\Qmat)=\cbra{i|Q_i\ne I}$, and the function $f$ depends on the measurement scheme. For different measurement schemes,  we show in the following different choices of $\Kcal(\Pmat)$ and the function $f$ that give an unbiased estimation
\begin{equation}\label{eq:unbiasedestimation}
    \mathbb{E}[\hat{\omat}] = \tr(\Omat \rho) 
\end{equation}
 where the average is over $\Kcal(\Pmat)$.  
 
An importance sampling method~\cite{wecker2015progress}, which is also called  $l_1$ sampling, corresponds to the case with $\Plmat= \Olmat$,  $\Kcal(\Pmat_l) = |\alpha_l|/\sum_l |\alpha_l|$, and $f(\Pmat, \Olmat,\Kcal) = \Kcal( \Pmat)^{-1} \delta_{\Pmat , \Olmat}$. Heuristic grouping methods, such as the one using largest degree first (LDF) grouping~\cite{kandala2017hardware,verteletskyi2020measurement,o2016scalable}, divide $\mathcal O = \{\Omat_l\}$ into several groups $\mathcal S_j$ such that  $\cup_j \mathcal S_j = \mathcal O$, $\mathcal S_j\cap \mathcal S_{j'}=\varnothing,\forall j\neq j'$.
For each group $\mathcal{S}_j$, measurement $\Pmat_j$ is assigned such that  $\Qmat \triangleright \Pmat_j,\forall \Qmat\in \mathcal S_j$ with probabilities $\Kcal(\Pmat_j)$ chosen either uniformly or based on the total weight of the observables in the set $\Pmat_j$. The function is chosen as $f(\Pmat_j, \Qmat,\Kcal) = \Kcal(\Pmat_j)^{-1} \delta_{\Qmat \in \mathcal{S}_j}$. \sun{
Although finding the optimal grouping strategy is NP-hard in decomposition of the complex Hamiltonian, the heuristic grouping methods could give sub-optimal strategies, especially for the Hamiltonian containing many compatible terms~\cite{verteletskyi2020measurement}.}

The conventional classical shadow (CS) method~\cite{huang2020predicting} considers the full-weight Pauli basis set $\Pmat \in  \{X,Y,Z\}^{\otimes n}$ with a uniform probability $\Kcal(\Pmat)=1/3^n$. One of its generalization is to consider locally biased classical shadow (LBCS)~\cite{hadfield2020measurements} with product and biased probability $\Kcal(\Pmat)=\prod_i \Kcal_i(P_i)$, where $\Kcal_i(P_i)$ represents the probability of measuring the $i$th site with the basis $P_i$. For the CS and LBCS methods, the function is defined as $f(\Pmat, \Qmat,\Kcal) = \prod_i f_i(P_i, Q_i,\Kcal_i)$ with $f_i(P_i, Q_i,\Kcal_i) = \delta_{Q_i, {I}_2} + \Kcal_i(P_i)^{-1} \delta_{Q_i, P_i}$. Huang \textit{et al}. further proposed the derandomized shadow method, in which the basis set is deterministically chosen by a {classical} greedy algorithm~\cite{huang2021efficient}. 
For the CS methods, the randomized measurement is implemented by applying random local Clifford unitaries. Besides the estimation of expected values of observables, we can also use classical shadows to calculate nonlinear properties of quantum states, in particular observables of higher state moments, as suggested in Refs.~\cite{huang2020predicting,elben2020mixed,neven2021symmetry,vitale2021symmetry}. We refer to Supplemental Materials~\cite{NoteX} for detailed discussions on the implementation of the CS scheme and the measurement cost complexity. 

Prior experiments have implemented the original CS method using uniform probability distribution. In particular, Struchalin \textit{et al.}~\cite{struchalin2021experimental} demonstrated the estimation of local observables and the state fidelity with uniformly random stablizer measurements on an optical system, and Elben \textit{et al.}~\cite{elben2020mixed} used prior experimental data of trapped ions from randomized measurements to detect the  bipartite entanglement. Here we focus on all the latest CS methods and compare them to other advanced measurement schemes. We consider the tasks of measuring linear and nonlinear observables and show the application and advantage of using classical shadows.

\begin{figure}[tbp]
	\includegraphics[scale=1]{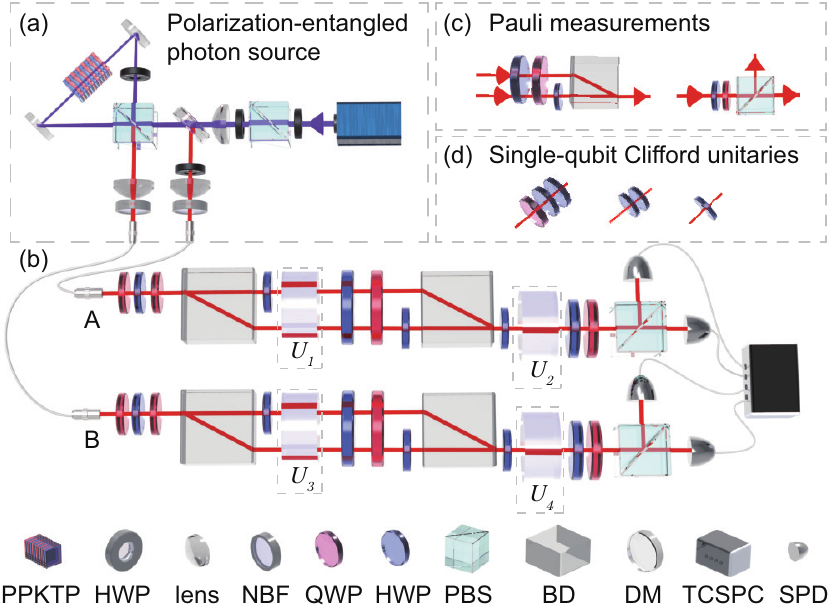}
	\caption{Schematic illustration of the experimental setup. (a)  The setup to generate maximally polarization-entangled photon pair. (b)  Two photons are sent into BD to generate a  four-qubit hyper-entangled state. (c)  Experimental setup to  implement the Pauli measurements. (d)  The single-qubit Clifford operations ($U_i$) are realized with different settings of waveplates. NBF: narrow-band filter. DM: dichroic mirror.}
	\label{Fig:Setup}
\end{figure}

\vspace{6pt}

\emph{{Experimental setup.}---}We implement the advanced measurement schemes on a photonic four-qubit Greenberger–Horne–Zeilinger (GHZ) state with ideal form of $\ket{\text{GHZ}_4} =(\ket{0000}+\ket{1111})/\sqrt{2}$. As shown in Fig.~\ref{Fig:Setup}(a), the polarization-entangled photons are generated from a periodically poled potassium titanyl phosphate (PPKTP) crystal in a Sagnac interferometer~\cite{Kim2006}, which is bidirectionally pumped by an ultraviolet (UV) laser diode with central wavelength of 405~nm. The two photons are entangled in the polarization degree of freedom (DOF) with ideal form of $\ket{\Phi^+}=(\ket{HH}+\ket{VV})/\sqrt{2}$, where $\ket{H}$ and $\ket{V}$ denote horizontal and vertical polarization, respectively. Each photon is then extended to its path DOF by passing through a beam displacer (BD) which transmits vertical component and deviates horizontal component. Thus, a four-qubit hyper-entangled state $\ket{\text{GHZ}_4}=(\ket{HhHh}+\ket{VvVv})/\sqrt{2}$ is generated, where $h$ and $v$ denote the path DOF. The qubit is encoded in the polarization DOF as $\ket{H(V)} \rightarrow \ket{0(1)}$, and in the path DOF as $\ket{h(v)} \rightarrow \ket{0(1)}$~\cite{Ding2020}. The measurements on basis $\Pmat$ on polarization DOF and path DOF are realized with setups shown in Fig.~\ref{Fig:Setup}(c). The single-qubit Clifford unitaries on either polarization or path DOF are realized by sets of half-wave plate (HWP) and quarter-wave plate (QWP) as shown in Fig.~\ref{Fig:Setup}(d). All the photons are collected with single-mode fibres and detected by single-photon detectors (SPD). The arriving time (time tag) of each photon is recorded by a time-correlated single-photon counting (TCSPC) system. By counting the time tags, the coincidence (as low as 1) on each measurement basis can be determined, as well as its corresponding statistical time. 
We implement the quantum state tomography (QST) on the prepared state $\rho_{\text{exp}}^{\text{GHZ}_4}$, and calculate that the fidelity of $\rho_{\text{exp}}^{\text{GHZ}_4}$ and $\ket{\text{GHZ}_4}$ is $\mathcal F=0.9546\pm0.0006$. We refer to~\cite{NoteX} for more details of experimental demonstrations and data processing.   
  

\begin{figure}[t]
	\includegraphics[width=1\linewidth]{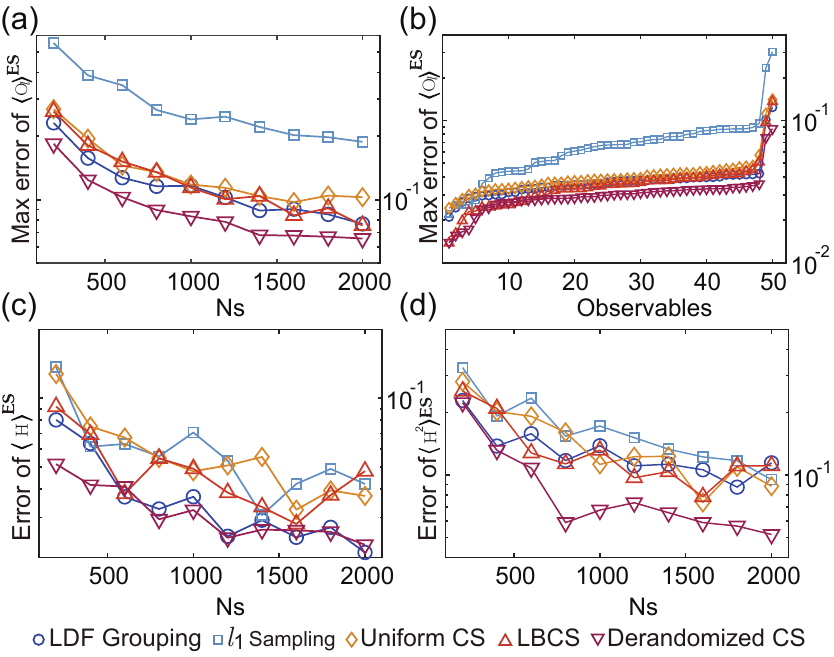}
	\caption{The error of observable estimations with five different measurement schemes. (a) The maximum absolute error of $\ES{\Olmat}$ over $50$ local observables $\Olmat$ that are randomly selected from the Pauli set with different number of $N_s$.
	(b) The maximum absolute error of $\ES{\Olmat}$ with different number of local observables, each of which we fix $N_s=2000$.
	(c) and (d) are the errors of estimated energy $\ES{H}$ and that of estimated Hamiltonian moment $\ES{H^2}$ with different $N_s$. In each measurement basis we collect five coincidences for data processing, and we run 20 independent repetitions of the entire setup.}
	\label{Fig:Error_vs_Observable}
\end{figure}

\emph{{Estimation of observables.}---}We perform the classical shadow (CS) schemes (uniform CS, locally biased CS and derandmized CS) as well as conventional schemes ($l_1$ sampling and LDF grouping) on the prepared state $\rho_{\text{exp}}^{\text{GHZ}_4}$. We randomly select $50$ observables $\Olmat$ ($l\leqslant 50$) that are tensor products of Pauli operators acting non-trivially on maximally two qubits.
\hl{The measurement bases are determined according to the target observables and the measurement scheme. Experimentally, the estimation is determined with the results from $N_s$ measurements (also called samples), in each of which we collect five coincidences.}
We post process the measurement results using Eq.~\eqref{eq:estimator_main} to estimate the expectation value of target observables.
{The maximal errors of estimated expectation $\ES{\Olmat}$ over 50 Pauli observables $\Olmat$ with five estimation schemes are shown in Fig.~\ref{Fig:Error_vs_Observable}, where the error is defined as the difference between $\ES{\Olmat}$ and results with direct measurement of $\Olmat$ on $\rho_{\text{exp}}^{\text{GHZ}_4}$.} In Fig.~\ref{Fig:Error_vs_Observable}(a), we observe that the maximal error of $\ES{\Olmat}$ decreases with {an increasing number of measurements} $N_s$. Except for $l_1$ sampling, we observe the maximum error is reduced to 0.1 when $N_s=2000$. Next, we fix $N_s$ and investigate the maximal error with an increasing number of observables, and the results are shown in Fig.~\ref{Fig:Error_vs_Observable}(b). We observe that the accuracy with the derandomized CS method outperforms those with other schemes, especially the $l_1$ sampling method.

Moreover, we demonstrate the estimation of energy and Hamiltonian moment in the variational quantum simulation~\cite{kokail2019self,cerezo2020variational,kowalski2020quantum,vallury2020quantum}.
We consider the Hamiltonian in the form of $H = J\sum_i (Z_{i} Z_{i+1} + X_{i} Y_{i+1} + Y_{i} Z_{i+1} + X_{i} Z_{i+1}) + h \sum_i X_i$ with the periodic boundary condition.
The results for $\ES{H}$ and its second-order moments $\ES{H^2}$ with normalized strength $J=h=1/4$ are shown in Fig.~\ref{Fig:Error_vs_Observable}(c) and Fig.~\ref{Fig:Error_vs_Observable}(d), respectively. Similarly, the error decreases with an increasing of $N_s$. We also observe that the results with LDF grouping and derandomized CS schemes outperform other schemes for the energy estimation as shown in Fig.~\ref{Fig:Error_vs_Observable}(c), while derandomized CS shows significant advantage in the estimation of $\ES{H^2}$ as reflected in Fig.~\ref{Fig:Error_vs_Observable}(d) owing to many large-support terms in $H^2$. One can expect that the advanced measurement schemes could be more competitive when the problem size increases. We leave the discussion on statistical errors, {numerical simulations with noiseless state, noise robustness, and results for cluster Hamiltonian with Ising interactions and the hydrogen molecular Hamiltonian} to Supplemental Materials~\cite{NoteX}.
\vspace{6pt}

\begin{figure}[t]
\includegraphics[width=1\linewidth]{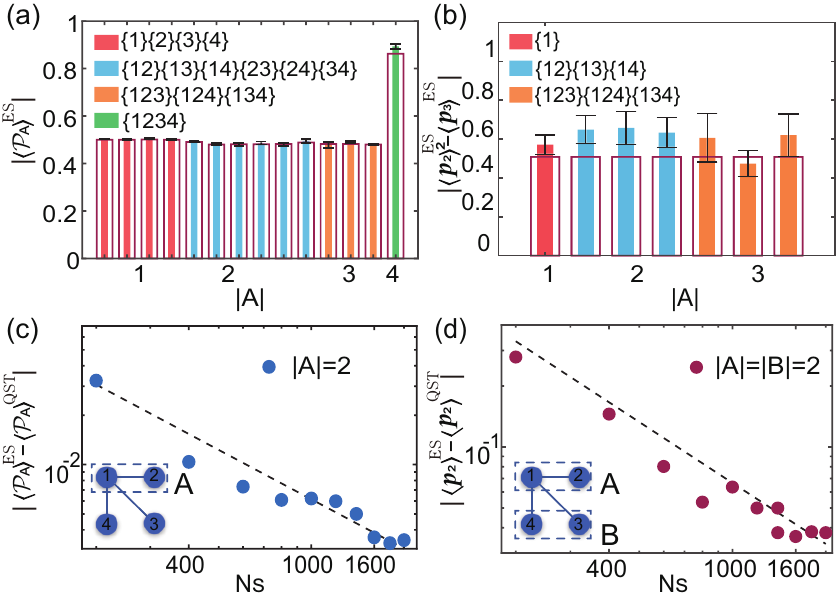}
\caption{Experimental results for estimating nonlinear functions with classical shadows.
(a) The estimation of subsystem purity $\ES{\mathcal P_A}$ with the different subsystems $A$. The colored bars represent the results from CS method, while the red sticks represent the results from QST {for comparison}. 
(b) The estimation of ${p_2^2}-{p_3}$ for different subsystem partitioning of the prepared GHZ state, which clearly shows the violation of the $p_3$-PPT condition. The number of measurements $N_s$ in (a) and (b) is fixed as $1000$ for the CS method, and the standard deviation is obtained by repeating the experiment 10 times. \zt{Here, we collect one coincidence at each measurement.}
The dots in (c) and (d)  represent the errors of $\ES{\mathcal P_A}$ and that of $\ES{ p_2}$ with different $N_s$, respectively. The dashed line represents the scaling of $\propto 1/N_s$. The $\ket{\text{GHZ}_4}$ is a specific graph state, corresponding to a star graph, which is exhibited in the insets. 
}
\label{Fig:subpurity}
\end{figure}

\emph{{Estimation of nonlinear function and entanglement structure.}---} We divide the four-qubit GHZ state $\rho_{\text{exp}}^{\text{GHZ}_4}$ into two subsystems $A$ and $B$, where $B$ is the complement set ($A \cup B = \{1,2,3,4\}$ and $A \cap B = \varnothing$).
Each subsystem contains $|A|$ and $|B|$ qubits, respectively. The purity of subsystem $A$ can be measured on two copies of $\rho_A$ by $\mathcal P_A=\tr[\rho_A^2] = \tr[\Pi_A \rho \otimes \rho]$ where $\Pi_A$ is the local swap operator of two copies of the subsystem $A$~\cite{RevModPhys.81.865,brydges2019probing}.
\hl{Instead, we can make use of the classical shadows to predict the expectation of high-order target functions.
The classical shadows of the underlying state $\rho$ can be generated by applying single-qubit Clifford unitary $U_i$ on the $i$th qubit drawn from a uniformly random distribution and collecting its corresponding outcome $\ket{\bmat_i}$ from projective measurements. The unbiased estimator $\hat{\rho}$  is given by $\hat{\rho}= \bigotimes_i (3U_i  \ket{ \bmat_i}\bra{ \bmat_i}U_i- I_2)$, and $\mathbb{E}[\hat{\rho}] =\rho$ (More details can be found in \cite{NoteX}).}
Note that the estimator of the subsystem state $A$ can be generated by choosing the index of qubit $i \in A$.
\zt{In experiment, we generated $N_s = 1000$ independent classical shadows $\hat\rho_A^{(k)}$ obtained from measurements on the prepared state under uniformly random local Clifford unitaries.}
Then, we randomly select two independent $\hat\rho_A^{(k_1)}$ and $\hat\rho_A^{(k_2)}$, and estimate the subsystem purity by {$\hat{\mathcal P}_A=\sum_{k_1 \neq k_2}\tr[\hat\rho_A^{(k_1)}\otimes\hat\rho_A^{(k_2)}]/N_s(N_s-1)$. Here, we improve the estimation accuracy by exploiting all the distinct samples~\cite{hoeffding1992class,elben2020mixed,huang2020predicting}}. 
Fig.~\ref{Fig:subpurity}(a) shows the estimation results for the subsystem purity estimation $\mathcal P_A$ for all possible divisions of subsystems. We observe that {$\ES{\mathcal P_A}<\ES{\mathcal P_{AB}}$} for all the subsystems $A \subseteq \{1,2,3,4\}$, which certifies genuine multipartite entanglement of the prepared state~\cite{RevModPhys.81.865}. 

We next demonstrate another entanglement detection based on the positive partial transpose (PPT) condition, which checks whether the partially transposed (PT) density matrix $\rho_{AB}^{T_A}$ has negative eigenvalues.
We consider the PT-moments $p_n = \tr\left[ (\rho_{AB}^{T_A})^n\right]$, and it has been shown that the state must be entangled if $p_2^2>p_3$~(see Refs.~\cite{elben2020mixed,neven2021symmetry}).
Note that $\tr\left[ (\rho_{AB}^{T_A})^n\right]=\tr\left[ \overrightarrow {\Pi}_A \overleftarrow {\Pi}_B \rho_{AB}^{\otimes n}\right]$, where $\overrightarrow {\Pi}_A $ and $\overleftarrow {\Pi}_B$ are $n$-copy cyclic permutation operators that act on the subsystems $A$ and $B$ respectively. 
The typical procedure to estimate $p_n$ requires measuring the observable $ \overrightarrow {\Pi}_A \overleftarrow {\Pi}_B $ on $n$ copies of quantum states. Instead, we can construct the $U$-statistic estimator of $p_n$ by summing over all possible pairs of the independent classical shadows~\cite{neven2021symmetry}:
$
\hat{p}_n=\frac{1}{ n!\binom{N_s}{n}}  \sum_{k_1 \neq \cdots \neq  k_n}  \tr \left[ \overrightarrow {\Pi}_A \overleftarrow {\Pi}_B \hat \rho_{AB}^{(k_1)} \otimes \cdots \otimes \hat \rho_{AB}^{(k_n)} \right]
$
, and the estimator is unbiased as $\mathbb{E}[\hat p_n]=p_n$.
The PT-moment can be efficiently computed as the summands are tensor products of local density matrix and is complete to factorize into contractions of single-qubit matrices~\cite{elben2020mixed}. 
The estimation of $\abs{(\ES{{p_2}})^2-\ES{{p_3}}}$ for different subsystem divisions are shown in Fig.~\ref{Fig:subpurity}(b)  which clearly violates the $p_3$-PPT condition ($p_2^2>p_3$) and indicates the genuine bipartite entanglement of the prepared state.
Compared to the purity condition ($\tr(\rho_A^2) < \tr(\rho_{AB}^2)$), $p_3$-PPT condition can be applied to detect entanglement of mixed state~\cite{elben2020mixed}, and more details can be found in~\cite{NoteX}. 

The estimation errors of subsystem purity $\abs{\ES{\mathcal P_A}-\QST{\mathcal P_A}}$ and the PT-moment~$\abs{\ES{ p_2}-\QST{p_2}}$ of the case $A=\{1,2\}$~and $B=\{3,4\}$ are shown in Fig.~\ref{Fig:subpurity}(c) and Fig.~\ref{Fig:subpurity}(d),~respectively, {where $\QST{\bullet}$ is the results calculated with reconstructed $\rho_{\text{exp}}^{\text{GHZ}_4}$} from quantum state tomography. Similarly, we observe that the estimation can be inferred using a small number of $N_s$ and becomes more accurate when $N_s$ increases. The estimation error decays proportionally to $1/N_s$ for a small number of samples, different from the asymptotic decay rate in the large sample limit. We also discuss the sample complexity for estimating general nonlinear function in the Supplemental Materials~\cite{NoteX}.

\emph{{Conclusion.}---}In this work, we experimentally study the feasibility of quantum measurements. We compare the advanced measurement schemes with no increase in the circuit depth, and show that the (derandomized) classical shadow method outperforms other advanced measurement schemes, especially the naive $l_1$ grouping measurement method, in estimating linear  observables, and it applies to extract the nonlinear functions of states. 
While we demonstrate the measurement on a small quantum device, these  measurement schemes works naturally for problems with larger sizes. Since the Hamiltonian of a larger problem could be even more complicated, the advanced measurement schemes could hence show more advantages in reducing the measurement cost.
Several other measurement schemes were posted very recently~\cite{wu2021overlapped,hadfield2021adaptive, hillmich2021decision}, which improve the energy estimation by introducing optimized  measurement schemes within the unified framework. The only difference is the selection of the measurement basis, and thus one can similarly compare those measurement schemes by experiments. {Also, we experimentally demonstrate that the classical shadow method applies to the estimation of Hamiltonian moments $\braket{H^n}$, which can be leveraged to correct the ground state energy obtained from the variational approach~\cite{vallury2020quantum,kowalski2020quantum} or in the adaptive variational quantum algorithms~\cite{grimsley2019adaptive,tang2019qubit,zhang2020lowdepth}. Moreover, one could minimize the variance of eigenenergy $\braket{H^2}-\braket{H}^2$ to  prepare the excited states of $H$ in the variational quantum simulation~\cite{kokail2019self}.} Those tasks generally require a prohibitively large number of measurements, which however could be significantly alleviated using classical shadows. We also demonstrate the detection of genuine entanglement using classical shadows, whose extension to general entanglement structure detection deserves future studies. 

Our work verifies the possibility of efficient  measurement of quantum states and paves the way for fast quantum processing using near-term quantum devices. {One direction for further research is to incorporate error mitigation into the measurement schemes~\cite{bravyi2020mitigating,Su2021errormitigationnear,li2017efficient,temme2017error,mcclean2020decoding,sun2020mitigating2,endo2020hybrid}.   Error mitigation is able to suppress device errors, which leads the estimation more accurate compared to the ideal value. As the CS scheme could be robust to shot noise, the combination of these two schemes is expected to make the estimation with high accuracy as well as high confidence.}

\begin{acknowledgments}
We thank Bujiao Wu and Vlatko Vedral for insightful discussions on the framework. This work is supported by the National Natural Science Foundation of China (Grant No. 11974213 and No. 92065112), National Key R$\&$D Program of China (Grant No. 2019YFA0308200), and Shandong Provincial Natural Science Foundation (Grant No. ZR2019MA001 and No. ZR2020JQ05), Taishan Scholar of Shandong Province (Grant No. tsqn202103013) and Shandong University Multidisciplinary Research and Innovation Team of Young Scholars (Grant No. 2020QNQT).

\end{acknowledgments}


%

\newpage
\clearpage
\widetext

\allowdisplaybreaks

\section*{Supplemental Material for ``Experimental quantum state measurement with classical shadows"}

\author{Ting Zhang}
\thanks{These two authors contributed equally }
\affiliation{School of Physics, Shandong University, Jinan 250100, China}

\author{Jinzhao Sun}
\thanks{These two authors contributed equally }
\affiliation{Center on Frontiers of Computing Studies, Peking University, Beijing 100871, China}
\affiliation{Clarendon Laboratory, University of Oxford, Parks Road, Oxford OX1 3PU, United Kingdom}

\author{Xiao-Xu Fang}
\affiliation{School of Physics, Shandong University, Jinan 250100, China}

\author{Xiao-Ming Zhang}
\affiliation{Center on Frontiers of Computing Studies, Peking University, Beijing 100871, China}
\affiliation{Department of Physics, City University of Hong Kong, Tat Chee Avenue, Kowloon, Hong Kong SAR, China}

\author{Xiao Yuan}
\email{xiaoyuan@pku.edu.cn}
\affiliation{Center on Frontiers of Computing Studies, Peking University, Beijing 100871, China}

\author{He Lu}
\email{luhe@sdu.edu.cn}
\affiliation{School of Physics, Shandong University, Jinan 250100, China}


\maketitle

\section{Methods}

\subsection{Framework for measuring quantum states}

We first review the unified framework for quantum state measurement with no increase in the circuit depth, recently proposed in Ref.~\cite{wu2021overlapped}.
We suppose that the target observable can be decomposed into the Pauli basis $\Omat = \sum_l \alpha_l \Olmat$ with $\Olmat \in \{I,X,Y,Z\}^{\otimes n}$.
Here,  we use the bold format to represent the $n$-qubit Pauli operators $\Olmat$ and the subscript $l$ of $\Olmat$ to represent the $l$th $n$-qubit Pauli operators in the decomposition. Without loss of generality, we denote an $n$-qubit  Pauli operator as $\Qmat = \otimes_{i=1}^n Q_i$ with $Q_i \in \{I, X,Y,Z\}$ being the single-qubit Pauli operator that acts on the $i$th qubit.

Provided the target observable and the measurement scheme, we first determine a measurement basis set $\{\Pmat\}$ and the corresponding  probability distribution $\{\Kcal\}$, and  then generate an estimation of $\tr(\rho \Omat)$ by measuring $\rho$ with $\Pmat$ selected  from the basis set over the distribution $\Kcal(\Pmat)$.
The estimator for the observable $\Omat$ with measurement $\Pmat$ is given by
\begin{equation}
\hat{\omat}(\Pmat) = \sum_l \alpha_l  f(\Pmat,\Olmat,\Kcal )\mu(\Pmat, \textrm{supp}(\Olmat)),
\label{eq:estimator_SM}
\end{equation}
{where $\mu(\Pmat,\textrm{supp}(\Olmat)) := \prod_{i\in \textrm{supp}}\mu(P_i)$ with $\mu(P_i)$ being the single-shot outcome by measuring the $i$th qubit of state $\rho$ with the Pauli basis $P_i$, and the support of $\Qmat$ defined by $\textrm{supp}(\Qmat):=\cbra{i|Q_i\ne I}$.} 
In the main text, we show the explicit forms of the probability distribution $\Kcal(\Pmat)$ and function $f$ for importance sampling, grouping and classical shadow algorithms, which give an unbiased estimation 
\begin{equation}
    \mathbb{E}[ \hat{\omat}] = \tr(\Omat \rho).
\end{equation}

The variance of the estimator in Eq.~\eqref{eq:estimator_SM} for a single sample could be calculated by the definition as
\begin{equation}
\textrm{Var}[\hat{\omat}] = \mathbb{E}_{\Pmat} \sum_{ l,l' } \alpha_{l} \alpha_{l'} \tr(\rho \Olmat \Olpmat)  f(\Pmat, \Olmat, \Kcal) f(\Pmat, \Olpmat, \Kcal) - \tr(\rho \Omat)^2,
\label{eq:Var_general}
\end{equation}
where we use the equality $\mathbb{E}_{\mu(\Pmat)} \mu(\Pmat, \operatorname{supp}(\Olmat)) \mu(\Pmat, \operatorname{supp}(\Olpmat))=\mathbb{E}_{\mu(\Pmat)} \mu(\Pmat, \operatorname{supp}(\Olmat\Olpmat) )=\tr(\rho \Olmat \Olpmat)$. The detailed proof can be found in Refs.~\cite{wu2021overlapped,hadfield2020measurements}.
In the following, we will discuss the relations of the  measurement algorithms within this framework.




\emph{Importance sampling}.  Importance sampling is also referred as the $l_1$ sampling. The measurement $\{\Pmat\}$ is selected as the observables $\{\Olmat\}$, and the corresponding probability  is determined by the weight of the observable as $\Kcal(\Pmat_l) = |\alpha_l|/\vabs{\bm \alpha}_{1}$. Here, $\vabs{\bm \alpha}_{1}$ is the $l_{1}$ norm of $\bm \alpha = \pbra{\alpha_{1}, \ldots, \alpha_{L}}$ as $\vabs{\bm \alpha}_{1} = \sum_{l =1}^{L} \abs{\alpha_{l}}$.
The function $f$ is defined by
\begin{equation}
    f_{l_1}(\Pmat, \Olmat,\Kcal) = \Kcal( \Pmat)^{-1} \delta_{\Pmat , \Olmat}.
\end{equation}
From Eq.~\eqref{eq:Var_general}, the variance of importance sampling   could be calculated by
\begin{equation}
    \textrm{Var}[\hat{\omat}] = \|\bm \alpha\|^2 - \tr(\rho \Omat)^2,
\end{equation}
which is directly related to the $l_1$ norm of the coefficients.

 
\emph{Grouping.} The essential idea for the grouping is that we first allocate observables $\Olmat$ to several non-overlapped sets, which satisfies that any {two observables $\Omat_{l}$ and  $\Omat_{l'}$ in each set are compatible with each other, i.e., $\Omat_l \triangleright \Omat_{l'}$ or $\Omat_{l'} \triangleright \Omat_{l}$.} 
Note that when the Pauli observables are
compatible with each other, their expectation values can be simultaneously obtained by measuring in one basis.
While finding  the  optimal measurement basis sets for the observables is NP-hard, several heuristic measurement basis have been proposed that runs in a polynomial time~\cite{verteletskyi2020measurement,wu2021overlapped}.
Here, we focus on the largest degree first (LDF) grouping method, whereas other grouping methods can be analyzed in a similar way.
We divide $\mathcal O = \{\Omat_l\}$ into several groups $\mathcal S_j$ such that  $\cup_j \mathcal S_j = \mathcal O$, $\mathcal S_j\cap \mathcal S_{j'}=\varnothing, \forall j\neq j'$.
For each group $\mathcal{S}_j$, measurement $\Pmat_j$ is assigned such that we can measure any observable $\Qmat$ in the $j$th set $\mathcal{S}_j$ with measurement $\Pmat_j$, i.e., $\Qmat\triangleright \Pmat_j,\forall\Qmat\in \mathcal S_j$ ($\Qmat$ element-wise commutes with $\Pmat_j$).
The  probability $\Kcal(\Pmat_j)$ can be chosen either uniformly or based on the total weight of the observables in the $j$th set, i.e. $\Kcal(\Pmat_j) = \vabs{\bm e_j}_1/\vabs{\alpha}_1$.
The function $f$ of the grouping method integrated with the importance sampling is chosen by 
\begin{equation}
    f_{\rm group}(\Pmat_j, \Qmat,\Kcal) = \Kcal(\Pmat_j)^{-1} \delta_{\Qmat \in \mathcal{S}_j}.
\end{equation}
From Eq.~\eqref{eq:Var_general}, the variance of the grouping method may be calculated by
\begin{equation}
\textrm{Var}[\hat{\omat}] = \sum_j \Kcal(\Pmat_j)^{-1} \sum_{ l,l': \Olmat,\Olpmat \in \mathcal{S}_j } \alpha_{l} \alpha_{l'} \tr(\rho \Olmat \Olpmat)  - \tr(\rho \Omat)^2.
\label{eq:var_group}
\end{equation}
This uses the definition of $f_{\rm group}$, which is nonzero only if $\Olmat \in \mathcal{S}_j$.

\emph{Classical shadow (CS)}.  We first perform randamized measurements on each qubit and then post process these  classical outcomes to estimate the target observables.  {The probability distribution $\Kcal_i(P_i)$ that performs Pauli measurement $P_i$ on $i$th qubit is independent on each site, and  therefore the probability distribution for one measurement $\Pmat$ is  a product of distribution on each site $\Kcal(\Pmat) = \prod_i \Kcal_i (P_i)$.}
The  uniform CS method consider a uniform distribution over the Pauli basis as {$\Kcal_i(P_i) = 1/3$}, which is irrespective of the target observables. 
In Ref.~\cite{hadfield2020measurements}, the authors proposed that the local probability distribution {$\Kcal_i$} could be optimized to reduce the number of samples, termed as locally biased classical shadow method.
The function is defined by 
\begin{equation}
    f_{\rm CS}(\Pmat, \Qmat,\Kcal) = \prod_i f_i(P_i, Q_i,\Kcal_i) 
\end{equation}
with $f_i(P_i, Q_i,\Kcal_i) = \delta_{Q_i, {I}_2} + \Kcal_i(P_i)^{-1} \delta_{Q_i, P_i}$.
Note that the variance for the  CS method can be bounded by 
\begin{align}
\begin{aligned}
\text{Var}[{\hat{\omat}}] &\leq \sum_{l,l'}\alpha_l \alpha_{l'} f_{\text{CS}}\pbra{\Olmat, \Olpmat, \Kcal}\tr\pbra{\rho \Olmat\Olpmat}   \leq 3^{\text{supp(\Omat)}}\pbra{\sum_{l=1}^L \alpha_l}^2 
\end{aligned}
\label{eq:CS_var}
\end{align}
with $\text{supp}(\Omat) := \max_{l} \text{ supp}(\Olmat)$.
From Eq.~\eqref{eq:CS_var},  the variance for the uniform CS method scales exponentially to the support of the target observable, and  the uniform CS method hence could be inefficient for the estimation of nonlocal observables with large support. 

\emph{Derandomized classical shadow}.
Huang \textit{et al.} further proposed the derandomized CS algorithms, in which the measurement basis {$\Pmat$} is deterministically selected.
The derandomization algorithm first assigns a collection of $N_s$ completely random $n$-qubit Pauli measurements, and iteratively identifies the measurement basis that minimizes the conditional expectation value over all remaining random measurement assignments. As argued in the Ref.~\cite{huang2021efficient}, the derandomized measurement procedure have no assurance to be globally optimal, or  close to optimal. Nevertheless, it shows significant performance for the realistic molecular Hamiltonian, ranging from $8$ to $16$ qubits.


We then discuss the variance for the derandomization algorithm.
Since the measurement bases are derandomized, the variance form in Eq.~\eqref{eq:Var_general} is inappropriate to evaluate the performance of derandomization.
Suppose we have determined the measurement basis set $\{\Pmat\}$.
Similarly to the grouping method, we denote $\mathcal{S}_j$ containing all $\Olmat$ hitted by $\Pmat_j$ (element-wise commute with $\Pmat_j$). We denote $s_l$ as the total number of
times $\Olmat$ is hitted. Let $N_j$ be the total number of times during the measurement of $\Pmat_j$,   ${t}_{l,j}$ be the outcome of $\Olmat$ is $+1$. The measurement outcome associated with the measurement $\Pmat_j$ for observable  $\Olmat$ is $\hat{o}_{l,j} = 2{t}_{l,j}/N_j - 1$. As such, the estimator of the expectation value of observable $\Omat$ can be expressed by
\begin{equation}
    \hat \omat = \sum_j \sum_{l:\Olmat \in \mathcal{S}_j } \frac{\alpha_l N_j}{s_l}   \hat{o}_{l,j}.
    \label{eq:estimate_derand}
\end{equation}
One can check that if  $N_j>0$ ($\forall j$), i.e., every observable is assigned at least one measurement basis (one sample), the estimation in Eq.~\eqref{eq:estimate_derand}
is unbiased.

The variance of the estimator  $\hat \omat $ 
 is given by
\begin{equation}
\textrm{Var}[\hat \omat ] = \sum_{j} N_{j} \sum_{l,l':\Olmat,\Olpmat \in \mathcal
S_{j}} \frac{1}{s_{l} s_{l'}} \alpha_{l} \alpha_{l'} \operatorname{Cov}\left(\hat{o}_{l,j}, \hat{o}_{l',j}\right).
\label{eq:derand}
\end{equation}
Here, we use the fact that measurement outcomes from different  $\Pmat_j$ are independent  since the measurements $\Pmat_j$ are independent on each other.
We also note that the outcomes $\hat{o}_{l,j}$ are correlated, so the variance in Eq.~\eqref{eq:derand} depends on the covariance  $\operatorname{Cov}\left(\hat{o}_{l,j}, \hat{o}_{l',j}\right)$.

Derandomization indeed utilizes the compatible properties of observables when measuring in the predetermined basis. More specifically, the measurement outcome with one measurement basis $\Pmat$ can be repeatedly used to calculate multiple expectation value of observable that is compatible with $\Pmat$. 
One can observe that once the measurement bases are determined, it could be regarded as a special grouping method. While differently, it allows the observables to be assigned in different groups.
In the following subsection, we can find its underlying relations to the grouping  method within the measurement framework  described by Eq.~\eqref{eq:estimator_SM}.

It is worth to mention that in the derandomized CS algorithms, the estimation could be biased since there exists some observables that might not be hit by any measurement in $\{\Pmat\}$. It thus introduces an initial error $\varepsilon_0 = |{\sum_{j:\Olmat \not \in \{\Pmat\}} \alpha_l \tr(\rho \Olmat) }|$, which indicates the biases to the expectation. More detailed discussion and numerical simulation can be found in Ref.~\cite{wu2021overlapped}


\emph{Other measurement algorithms}.
Several other relevant works that do not introduce entangling gates for measurements were posted very recently~\cite{wu2021overlapped,hadfield2021adaptive, hillmich2021decision}.
These measurement schemes improve the  performance of energy estimation by introducing the optimized  measurement  basis and probability distribution. Note that the measurement basis could be deterministically selected given a certain number of measurements.
Wu \textit{et al.} proposed the overlapped grouping method that exploits the spirit of Pauli grouping and classical shadows. The numerical simulation shows significant improvement over the prior works~\cite{wu2021overlapped}.
Hadfield \textit{et al.}~\cite{hadfield2021adaptive} proposed an adaptive Pauli shadow algorithm to generate an estimation, and Hillmich \textit{et al.}~\cite{hillmich2021decision} proposed a decision diagrams method to generate an estimation.
It is worth noting that these methods are within the unified framework introduced in the main text. 
One may check the performance of different methods by the variance  as
\begin{equation}
    \mathrm{Var}[\hat \omat]=\sum_{j,k} \alpha_l\alpha_{l'} g(\Olmat,\Olpmat)\tr(\rho \Olmat\Olpmat) - \tr(\rho \Omat)^2
    \label{eq:VarOGM}
\end{equation}
where
$g(\Olmat,\Olpmat) = \left(\sum_{\Pmat:\Olmat \triangleright \Pmat} \Kcal(\Pmat) \right) ^{-1} \left(\sum_{\Pmat:\Olpmat \triangleright \Pmat} \Kcal(\Pmat) \right) ^{-1} {\sum_{\Pmat: \Olmat\triangleright \Pmat \land \Olpmat \triangleright \Pmat} \Kcal(\Pmat)}$.
Here, we use the defination in Eq.~\eqref{eq:estimator_SM}, and the denominator in Eq.~\eqref{eq:VarOGM} indeed represents the probability that the observable $\Olmat$ ($\Olpmat$) is effectively measured with the measurement basis $\Pmat$ (See Lemma 1 in Ref.~\cite{wu2021overlapped}).
One can check that it reduces the conventional grouping method  in Eq.~\eqref{eq:var_group} if we restrict that each observable can only be assigned into one group. Moreover, if the measurement bases are deterministically selected from the probability distribution and then fixed, it has the same spirit as that in Eq.~\eqref{eq:derand}.
We can similarly compare the performance of these methods using the experimental data and the corresponding post-processing method.

In this work, we experimentally demonstrate the estimation of multiple local observables and energy estimation. 
We also show that the measurement schemes can be naturally applied to estimate the Hamiltonian powers  $\braket{H^n}$, which can be  used to estimate the energy variance or to correct the ground state energy.
The higher moments of Hamiltonian generally comprises many terms, which might be challenging if we  measure each term directly. These advanced measurement schemes can be employed to save the number of measurements. Therefore, our results could be useful for the ground state energy estimation with near-term quantum devices, and could show more advantages when the system size increases larger.

\subsection{Classical shadows}
\label{appendix:CS}

As analyzed in the above section, in the CS method, we extract  the  properties  of  the quantum  state by performing randomized  measurements, which projects the quantum state to classical information over a properly chosen distribution. We can estimate other properties of the quantum state along this line.
In this section, we review the CS method proposed in Ref.~\cite{huang2020predicting}, and  discuss the estimation of  the  nonlinear  properties  of  quantum  state using the CS method.

Shadow tomography was first proposed by Aaronson~\cite{aaronson2019shadow}, and later Huang \textit{et al.} has showed that  one can predict multiple physical properties of quantum states with asymptotic scaling up to polylogarithmic factors.
The key ingredient of the CS algorithm is that one perform  random unitary operations $U$ to the quantum state, and  measure the rotated state $U\rho U^{\dagger}$ in the computation basis ${\bmat} \in \{0,1\}^{\otimes n}$. 
Making use of the classical outcomes $\ket{\bmat}$, one can reconstruct the unknown quantum state as 
\begin{equation}
    \rho = \mathcal{M}^{-1}(U^{\dagger}\ket{\bmat}\bra{\bmat}U) 
\end{equation}
where $\mathcal{M}$ is a quantum channel that depends on the ensemble of random unitary transformation.
One can prove that $\mathcal{M}$ is a depolarizing channel, and thus the explicit form of the inverted channel $\mathcal{M}^{-1}$ is $\mathcal{M}^{-1}_n(\rho) = (2^n+1)\rho - I_{2^n}$ for global Clifford operations $\rm{Cl}_{2^n}$ and $\mathcal{M}^{-1} = \otimes_n \mathcal{M}^{-1}_1$ for local Clifford operations $\rm{Cl}_2$, respectively.
We can investigate multiple properties of the quantum state by appropriately post processing the classical information obtained from the results measured on a single copy of quantum state.

In the experiment, we apply random local Clifford operations drawn from a uniform distribution, and perform projective measurements on the GHZ state to obtain the classical measurement outcome $\ket{\bmat}$.
Given the $k$th measurement outcome string $\bmat^{(k)}$, we can construct the classical shadow of the quantum state by
\begin{equation}
\hat{\rho}^{(k)}= \bigotimes_i (3U_i^{(k)\dagger} \ket{ \bmat_i^{(k)}}\bra{ \bmat_i^{(k)}}U_i ^{(k)}- I_2).
\label{eq:shadow_k}
\end{equation}
The unknown  quantum state can be estimated by averaging over all unitaries configuration sampled from a unitary $3$-design by $\hat{\rho}=\frac{1}{N_s}\sum_{k=1}^{Ns} \hat{\rho}^{(k)}$, which produces the exact state in expectation $\mathbb{E}[\hat \rho] =\rho$. Each copy of $\hat{\rho}$ can be regarded the classical shadow of the underlying quantum state, and hence we refer it as a classical shadow in this context. It is also called snapshot in the literature.
In practice, for each set of the applied unitary operations, the measurement could be repeated $N_r$ times to improve the experiment statistics.

In the task of observable estimation, we estimate the expectation value of $L$ local observables $\tr{ [\Omat_l\rho]}$, $l \leq L$.  Suppose the observables acting non-trivially on maximally $m$ qubits. The expectation values of  local observables can be efficiently calculated using the reduced density matrix.  From Eq.~\eqref{eq:CS_var}, $\mathcal{O}({3^m\log(L)}/{\varepsilon^2})$ samples suffices to predict $L$ arbitrary observables  $O_1$... $O_L$ up to additive error $\varepsilon$, where $m$ is the largest support of observables.  

For the original algorithm proposed in Ref.~\cite{huang2020predicting}, the authors used the median-of-means estimator to preclude the outlier corruption. Nevertheless, this median evaluation can be omitted in the  large samples limit $N_s \rightarrow \infty$. 
In the asymptotic limit $N_s \rightarrow \infty$, the estimator {$\hat\omat_l$} for $l$th observables obeys the normal distribution {$\hat \omat_l\sim \mathcal{N}(   \mathbb{E}[ \hat\omat_l], \textrm{Var}[\hat \omat_l]/N_s)$}. The failure probability can be calculated by
\begin{equation}
\textrm{Pr}[|\hat{\omat} - {\mathbb{E} [\hat\omat] }| > \varepsilon] \leq L \max_l \textrm{Pr}[|\hat{\omat}_l -{ \mathbb{E} [\hat\omat_l ]| }> \varepsilon] \leq L \exp \left( - \frac{N_s \varepsilon^2}{2 \max_l \textrm{Var}[\hat \omat_l]} \right).
\end{equation}
Therefore, the number of samples can be chosen by 
\begin{equation}
N_s \geq 2\log(L) \log(1/\delta) \max_l\textrm{Var}[\hat \omat_l] /\varepsilon^2
\end{equation}
such that the estimator obeys the failure probability within $\delta$ as $\textrm{Pr}[|\hat{\omat} - {\mathbb{E} [\hat\omat]|} > \varepsilon] \leq \delta$.
In both the numerics and experiments, we find that the median estimators performs consistent and robust against outliers.
In our experiments, we did not observe the advantage using the median evaluation, which is consistent with the results in Ref.~\cite{struchalin2021experimental}.  

As shown in Ref.~\cite{huang2020predicting}, the CS scheme can be naturally extended to estimate the  nonlinear properties of  quantum  states,  such as  observables  of  higher state moments, which can be expressed  as a linear function in the tensor product of multiple copies:  $\tr[O \rho \otimes \cdots  \otimes  \rho]$. Here, $O$ acts on multiple copies of the quantum state. 
For example, the second-order subsystem Renyi entropy can be written as $\tr[\rho_A^2] = \tr[\Pi_A \rho \otimes \rho]$, where $\Pi_A$ is the local swap operator of two copies of the subsystem $A$.
We also note that it is related to the second order of PT-moments by $\tr[\rho_{AB}^2]= \tr[\rho_{AB}^{T_A}\rho_{AB}^{T_A}]$.
To estimate the nonlinear function, we can perform joint measurements on multiple copies of the quantum state. While it might achieve lower sample complexity, it could be challenging to implement in experiments. In the following, we show the estimation from the measurements on a single-copy of the quantum state, following the discussions in Ref.~\cite{huang2020predicting} closely.

Suppose we have collected $N_s$ copies of the classical shadows {$\hat \rho_j$} and aim to estimate $\tr[O \rho \otimes \cdots  \otimes  \rho]$ using these classical shadows.
The estimator for the $j_k$th copy is constructed by 
\begin{equation}
\hat \rho^{(j_k)} = \bigotimes_{i=1}^N \left( 3U_i ^{(j_k) \dagger} \ket{ {\mathbf{b}_i}^{(j_k)}} \bra{ {\mathbf{b}_i}^{(j_k)}} U_i ^{(j_k)  }    - I_2, \right)
\label{eq:shadow_k2}
\end{equation}
where  $2$-design property of Clifford groups is used to get the explicit form.
We can estimate {$\tr[O \rho \otimes \cdots  \otimes  \rho]$ by $\hat o =\tr[O  \hat \rho_{j_1} \otimes \hat \rho_{j_2} \cdots \otimes \hat \rho_{j_n}]   $}, which produces the exact value in expectation as 
\begin{equation}
\mathbb{E}[\hat o] =\tr[O  \mathbb{E}\hat \rho_{j_1} \otimes \mathbb{E} \hat \rho_{j_2} \cdots \otimes \mathbb{E} \hat \rho_{j_n}]   = \tr[O \rho \otimes \cdots  \otimes  \rho]. 
\end{equation}
Here, we use the subscript $j$ to denote the $j$th copy,  and  abbreviate the classical shadow $\rho^{(j)}$ as $\rho_j$ when there is no confusion.

According to Born's rule, the estimation for nonlinear function is now given by

\begin{equation}
\begin{aligned}
\mathbb{E}[\hat o] &= \sum_{\mathbf{b}_{j_1}\mathbf{b}_{j_2}\cdots \mathbf{b}_{j_n}} \textrm{Pr}[ {\mathbf{b}} = \mathbf{b}_{j_1}\mathbf{b}_{j_2}\cdots \mathbf{b}_{j_n}] \tr[O  \hat \rho_{j_1} \otimes \hat \rho_{j_2} \cdots \otimes \hat \rho_{j_n}],  \\
\end{aligned}
\end{equation}
where $\textrm{Pr}[ {\mathbf{b}} = \mathbf{b}_{j_1}\mathbf{b}_{j_2}\cdots \mathbf{b}_{j_n}]  = \braket{\mathbf{b}_{j_1}|U_{j_1} \rho_{j_1} U_{j_1} ^ {\dagger}|\mathbf{b}_{j_1}} \braket{\mathbf{b}_{j_2}|U_{j_2} \rho_{j_2} U_{j_2} ^ {\dagger}|\mathbf{b}_{j_2}} \cdots \braket{\mathbf{b}_{j_n}|U_{j_n} \rho_{j_n} U_{j_n} ^ {\dagger}|\mathbf{b}_{j_n}}$ is the joint probability for the measurement outcomes $\mathbf{b}_{j_1}\mathbf{b}_{j_2}\cdots \mathbf{b}_{j_n} $, ($\mathbf{b}_{j_k} \in \{0,1\}^N$).
Given $N_s$ copies of measurement outcomes, we can estimate  $\tr[O \rho \otimes \cdots  \otimes  \rho]$ with classical computational complexity scaling as $\mathcal{O}(N_s^n  (nN)^2/ \log nN)$.

Under this scenario, we can estimate subsystem purity and the moments of the partially transposed density matrix, which could be used to quantify entanglement of the subsystems. The moments of partially transposed density matrix is $p_n=\tr[(\rho_{AB}^{T_A})^n]$, where $A$ and $B$ are the subsystems. Note the fact that $\tr\left[ (\rho_{AB}^{T_A})^n\right]=\tr\left[ \overrightarrow {\Pi}_A \overleftarrow {\Pi}_B \rho_{AB}^{\otimes n}\right]$, where $\overrightarrow {\Pi}_A $ and $\overleftarrow {\Pi}_B$ are $n$-copy cyclic permutation operators that act on the subsystems $A$ and $B$ respectively.   This evaluation requires to measure the observable $ \overrightarrow {\Pi}_A \overleftarrow {\Pi}_B $ on  $n$ copies of quantum states.  Instead, we can construct the unbiased estimator of $p_n$  by summing over all the distinct pairs of the independent classical shadows
\begin{equation}
\hat{p}_n=\frac{1}{\binom{N_s}{n} n!}  \sum_{k_1 \neq \cdots \neq  k_n} \tr \left[ \overrightarrow {\Pi}_A \overleftarrow {\Pi}_B \hat \rho_{AB}^{(k_1)} \otimes \cdots \otimes \hat \rho_{AB}^{(k_n)} \right],
\label{eqn:pn_1}
\end{equation} with the classical shadow $\hat \rho_{AB}^{(k_j)}$ ($k_j = 1, ..., n$) defined in Eq.~(\ref{eq:shadow_k2}).
Here, we have used the U-statistics estimator to improve the estimation accuracy, which replaces the multicopy state $\otimes_n \rho $ by a symmetric tensor product of multiple distinct shadows $\hat \rho_{AB}^{(k_j)}$~\cite{hoeffding1992class}.

From Eq.~(\ref{eq:shadow_k2}),  the summands in Eq.~(\ref{eqn:pn_1}) are tensor products of single-qubit density matrix, it is hence straightforward to calculate the PT moments by 
\begin{equation}
\hat{p}_n=\frac{1}{\binom{N_s}{n} n!}  \sum_{k_1 \neq \cdots \neq  k_n}\prod_{j\in A} \tr \left[ \hat \rho_j^{(k_1) T} \cdots \hat\rho_j^{(k_n) T} \right] \prod_{j\in B} \tr \left[\hat \rho_j^{(k_1) } \cdots \hat\rho_j^{(k_n) } \right].
\label{eqn:pn}
\end{equation}
Given $N_s$ measurement outcomes $\hat b_k$, the classical storage scales as $N_s^n |AB|$, and we can similarly use the stabilizer formalism to estimate $\hat p_n$, scaling as $\mathcal{O}(N_s^n (n|AB|)^2/ \log(n|AB|))$ instead of post-processing exponentially large matrix $\rho_{AB}$.  However, we should note that the variance of the estimator scales exponentially in the (sub)system size, as discussed in the next subsection.

\subsection{Error analysis for higher order nonlinear function}

In this section, we discuss the sample complexity to achieve the estimation of nonlinear function up to a certain error $\varepsilon$.
As reviewed in Sec.~\ref{appendix:CS}, to estimate the nonlinear function in $\rho$, for example $p_n = \tr((\rho^{T_A})^n)$, we first represent it as a linear function on the tensor product of the quantum state as 
$
    o = \tr(O  \rho \otimes \rho \cdots\otimes \rho  ) 
$
with $O$ acting on multiple copies.
To improve the estimation accuracy,
the estimation can be replaced by a symmetric tensor products of multiple distinct classical shadows 
$
\hat o =    \frac{1}{n!}   \sum_{\pi \in \mathcal{S}_n} \tr \left( O \hat  \rho_{\pi(i_1)} \otimes  \hat \rho_{\pi(i_2)} \otimes  \cdots  \otimes \hat \rho_{\pi(i_n)} \right)
$
Here,  $\mathcal{S}$ denotes the permutation group. Suppose we have collected { $N_s$ classical shadows} obtained from randomized measurements, and then we can improve the statistics by averaging all distinct pairs as 
\begin{equation}
\hat o =    \frac{1}{n! \binom{N_s}{n}}  \sum_{i_1 <i_2< \cdots <i_n } \sum_{\pi \in \mathcal{S}_k} \tr \left( O \hat  \rho_{\pi(i_1)} \otimes  \hat \rho_{\pi(i_2)} \otimes  \cdots  \otimes \hat \rho_{\pi(i_k)} )) \right).
\end{equation}




For a subsystem $AB$, the single-shot variance of the estimation $\tr(O \rho_{AB})$ can be bounded by
\begin{equation}
\textrm{Var}[ \tr (O \hat \rho )] \leq 2^{n|AB|}\tr(O^2) \leq 4^{n|AB|}\|O\|_{\infty}^2,
\label{eq:var_Orho}
\end{equation}
where $\|\cdot\|_{\infty}$ is the spectral norm (See Proposition 3 in Ref.~\cite{huang2020predicting}).
By the Chebyshev inequality,  when the number of samples $N_s$ satisfy $N_s\geq \text{Var}[\hat{o}]/\delta\varepsilon^2$, we achieve $\Pr\sbra{\abs{\hat{o} -  \tr (O \hat \rho )}\geq \varepsilon}\leq \delta$ with error $\varepsilon\ge 0$ and failure probability $\delta\in[0,1]$.
Therefore, the number of samples 
\begin{equation}
    N_s \geq 2^{n|AB|} \tr(O^2)/\delta\varepsilon^2
    \label{SMeq:sample_p2}
\end{equation}
suffices to achieve the estimation error $\varepsilon$ with failure probability $\delta$.
The inequality in Eq.~\eqref{SMeq:sample_p2} indicates the required number of measurements scales exponentially to the order and the size of the target system.

In Refs.~\cite{huang2020predicting,elben2020mixed}, it has been shown that the  variance for the second order function, for example, is related to two parts, including the variance $\textrm{Var}[\tr(O \hat \rho_{AB}^{(i)} \otimes \hat \rho_{AB}^{(j)} )]$, and the linear variance terms
$\textrm{Var}[\tr(O \hat \rho_{AB}^{(i)} \otimes \rho_{AB}]$ and $\textrm{Var}[\tr(O \rho_{AB} \otimes \hat \rho_{AB}^{(i)} ] $.
The former variance can be regarded as a classical shadow of the joint quantum state on two copies $\rho_{AB} \otimes \rho_{AB}$. From Eq.~\eqref{eq:var_Orho}, it is bounded by  $\textrm{Var}[\tr(O \hat \rho_{AB}^{(i)} \otimes \hat \rho_{AB}^{(j)} )] \leq 4^{|AB|} \tr(O^2)$. In the case of  second order of PT-moments, the operator is the local swap operator $O = \Pi_{AB}$, which satisfies $\Pi_{AB}^2 = I_2$.
The linear variance terms are $\textrm{Var}[\tr (\tilde O \hat \rho_i)]$ with $\tilde O = \rho_{AB}$, which is bounded by $\textrm{Var}[\tr (\tilde O \hat \rho_i)) \leq 2^{|AB|} \tr(\rho_{i}^2)$. Here we follow the convention in Sec.~\ref{appendix:CS} that abbreviates the independent shadow $\rho_{AB}^{(i)}$ as $\rho_i$. By counting all possible distinct pairs, one has
\begin{equation}
\begin{aligned} 
\operatorname{Var}\left[\hat{p}_{2}\right]   \leq   \frac{4}{N_s}  \operatorname{Var}[\tr(\rho_{A B} \hat{\rho}_{A B}) ]  + \frac{4}{N_s^2}  \operatorname{Var}\left[\tr\left(\Pi_{A B} \hat{\rho}_{A B}^{(1)} \otimes \hat{\rho}_{A B}^{(2)})\right] \right)    
\leq \frac{p_{2} 2^{|A B|+2} }{N_s  }+\frac{ 4^{|A B|+1}}{N_s^{2}}.
\end{aligned}
\end{equation}
In the large sample limit, the first terms dominates and thus the error decays proportionally to $1/\sqrt{N_s}$. When in the intermediate sample $N_s$ and the inequality holds $4^{|A B|} > 2p_{2} 2^{|A B|}$, the error decays proportionally to $1/{N_s}$, and the error decays proportionally to $1/\sqrt{N_s}$ in the large sample limit. More details can be found in Appendix D in Ref.~\cite{elben2020mixed}.

The general variance for  $n$th order functions can be analyzed similarly, including the variance of the joint quantum state $\textrm{Var}[\tr(O \otimes_i \hat \rho_{i} ) ) $ and the contribution from the lower order variance.
In the case of $n$th order PT-moments, the latter variance  has the  form   of
\begin{equation}
\textrm{Var} \left[ \sum_{k=1}^n C_k \tr((\rho^{T_A})^{n-k} \sum_{\pi \in \mathcal{S}_k} \hat  \rho_{\pi(i_1)}^{T_A} \hat \rho_{\pi(i_2)}^{T_A} \cdots \hat \rho_{\pi(i_k)}^{T_A}) \right],
\end{equation}
where $C_k$ represents the counting number for all the $k$th order configuration, and $\mathcal{S}$ denotes the permutation group. Here, we use the cyclic properties of trace.
Note that both terms of the variance can be transformed into the canonical form as  $\hat S = \textrm{Var}[\tr (\tilde O \otimes_i \hat \rho_i)]$, with $\tilde O$ is the operator acting on the subsystems. We can therefore iteratively compute the variance iteratively by counting all the contribution $\hat S$ from the $n$th order to the linear one.
For example, the estimation of $p_3$ in different regime of sample sizes $N_s$ was discussed in Ref.~\cite{elben2020mixed}.

In the above analysis, we approximate the multiple copies $\rho \otimes \rho \otimes \cdots \otimes \rho$ by a symmetric tensor product of $N_s$ independent shadows $\frac{1}{n!}\sum_{\pi \in \mathcal{S}_n} \hat \rho_{\pi(i_1)} \hat \rho_{\pi(i_2)} \cdots \hat \rho_{\pi(i_n)}$. 
In our experiments, as the system size is compatible for the current computing power, one may directly store the classical shadow of the full density matrix and compute the higher-order nonlinear functions as an alternative method. Note that this direct calculation is not scalable for large systems.

\subsection{Derivation of the PT-moments}
In this section, we prove the equality 
\begin{equation}
\tr\left[ (\rho_{AB}^{T_A})^n\right]=\tr\left[ \overrightarrow {\Pi}_A \overleftarrow {\Pi}_B \rho_{AB}^{\otimes n}\right]
\end{equation}
by definition.
L.H.S reads
\begin{equation}
\begin{aligned}
\tr\left[ (\rho_{AB}^{T_A})^n\right] &=
\tr\left[   \sum_{i_1,j_1,i_1',j_1' } \sum_{i_2,j_2,i_2',j_2' }  \cdots \sum_{i_n,j_n,i_n',j_n' }  \rho_{i_1'j_1,i_1'j_1'} \ket{i_1j_1} \bra{i_1'j_1'} \rho_{i_2'j_2,i_2j_2'} \ket{i_2j_2} \bra{i_2'j_2'} \cdots \rho_{i_n'j_n,i_nj_n'} \ket{i_nj_n} \bra{i_n'j_n'}\right]\\
&=   \sum_{i_1,j_1,i_1',j_1' } \sum_{i_2,j_2,i_2',j_2' }  \cdots \sum_{i_n,j_n,i_n',j_n' }   \rho_{i_1'j_1,i_1'j_1'} \rho_{i_2'j_2,i_2j_2'} \cdots \rho_{i_n'j_n,i_nj_n'}  \braket{i_n'j_n'|i_1j_1}\braket{i_1'j_1'|i_2j_2} \cdots \braket{i_{n-1}'j_{n-1}'|i_nj_n}\\
&=   \sum_{i_1,j_1,i_2,j_2,..., i_n,j_n }   \rho_{i_2j_1,i_1j_2} \rho_{i_3j_2,i_2j_3} \cdots \rho_{i_1j_n,i_nj_1}
\end{aligned}
\end{equation}R.H.S reads
\begin{equation}
\begin{aligned}
\tr\left[ \overrightarrow {\Pi}_A \overleftarrow {\Pi}_B \rho_{AB}^{\otimes n}\right] &=
\tr\left[   \sum_{i_1,j_1,i_1',j_1' } \sum_{i_2,j_2,i_2',j_2' }  \cdots \sum_{i_n,j_n,i_n',j_n' }  
\rho_{i_1j_1,i_1'j_1'} \ket{i_nj_2} \bra{i_1'j_1'}  \otimes \rho_{i_2j_2,i_2'j_2'} \ket{i_1j_3} \bra{i_2'j_2'}  \otimes  \cdots \otimes \rho_{i_nj_n,i_n'j_n'} \ket{i_{n-1}j_1} \bra{i_n'j_n'}\right]\\
&=  \sum_{i_1,j_1,i_1',j_1' } \sum_{i_2,j_2,i_2',j_2' }  \cdots \sum_{i_n,j_n,i_n',j_n' }  
\rho_{i_1j_1,i_1'j_1'}   \rho_{i_2j_2,i_2'j_2'}  \cdots \rho_{i_nj_n,i_n'j_n'} \delta_{i_n,i_1'} \delta_{i_1,i_2'} ... \delta_{i_{n-1},i_n'} \delta_{j_2,j_1'} \delta_{j_3,j_2'}...\delta_{j_n,j_{n-1}'}  \delta_{j_1,j_{n}'}  \\
&=   \sum_{i_1,j_1,i_2,j_2,..., i_n,j_n }   \rho_{i_1j_1,i_nj_2}\rho_{i_2j_2,i_1j_3} ... \rho_{i_nj_n,i_{n-1}j_1}\\
&=   \sum_{i_1,j_1,i_2,j_2,..., i_n,j_n }   \rho_{i_2j_1,i_1j_2} \rho_{i_3j_2,i_2j_3} \cdots \rho_{i_1j_n,i_nj_1}
\end{aligned}
\end{equation}
The last equality holds by replacing $k\rightarrow k+1 ~(\text{mod}~ n) ~k=1...n$, and we hence complete the proof.
In Ref.~\cite{elben2020mixed}, several useful equations are proven using the tensor network diagrams.

\section{Experimental implementation}
Before explaining our experimental procedure in detail, we first introduce the important optical components in the following~\cite{Li2021}:
\begin{itemize}
\item[1)] We use half-wave plates ($\rm{HWPs@\theta}$) and quarter-wave plates ($\rm{QWPs@\vartheta}$) to complete the unitary transformations. The $\rm{\theta}$ or $\rm{\vartheta}$ here refers to the angle between the fast axis of the waveplate and the vertical polarisation direction. The unitary transformations of waveplates acting on a quantum state can be expressed by Eq.~(\ref{eq:waveplates}).
\begin{align}\label{eq:waveplates}
    U_{\text{HWP}}=-\left( \begin{matrix}
	\cos 2\theta&		\sin 2\theta\\
	\sin 2\theta&		-\cos 2\theta\\
\end{matrix} \right),\, U_{\text{QWP}}=\frac{1}{\sqrt{2}}\left( \begin{matrix}
	1+i\cos 2\vartheta&		i\sin 2\vartheta\\
	i\sin 2\vartheta&		1-i\cos 2\vartheta\\
\end{matrix} \right)
\end{align}

\item[2)] A polarization beam splitter (PBS) has the function of transmitting photons in the direction of horizontal polarization but reflecting  photons in the direction of vertical polarisation.

\item[3)] A beam displacer (BD) is capable of fully transmitting vertically polarised photons, but deflecting them from their original path (about $\rm{3mm}$ in our experiment) when transmitting horizontally polarized photons.
\end{itemize}
\subsection{Polarization-entangled photon source}
The pump light is generated from an ultraviolet (UV) laser diode with central wavelength of 405~nm and full-width at half-maximum (FWHM) of 0.012~nm. The power intensity of pump light is adjusted by a HWP and a PBS. Then, the polarization of pump light is converted from $\ket{H_p}$ to $\ket{+_p}=\frac{1}{\sqrt{2}}\left( \ket{H_p} +\ket{V_p}\right)$ by a HWP set at $22.5^\circ$. 
\begin{figure}[h!tbp]
\includegraphics[width=0.41\columnwidth]{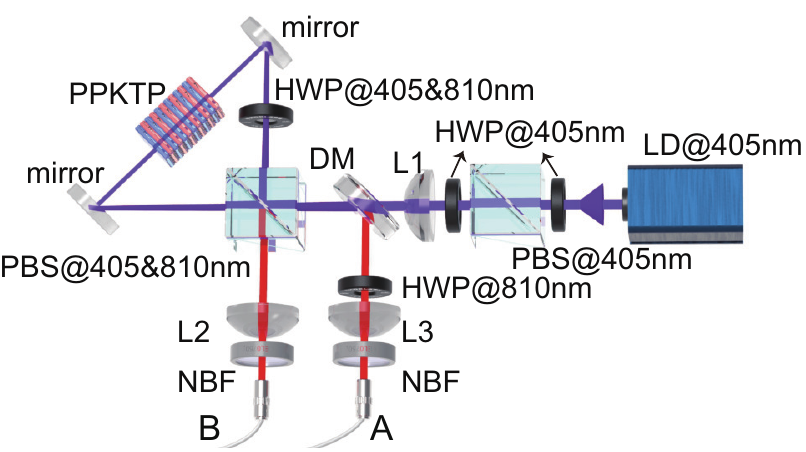}\hspace{1cm}
\includegraphics[width=0.35\columnwidth]{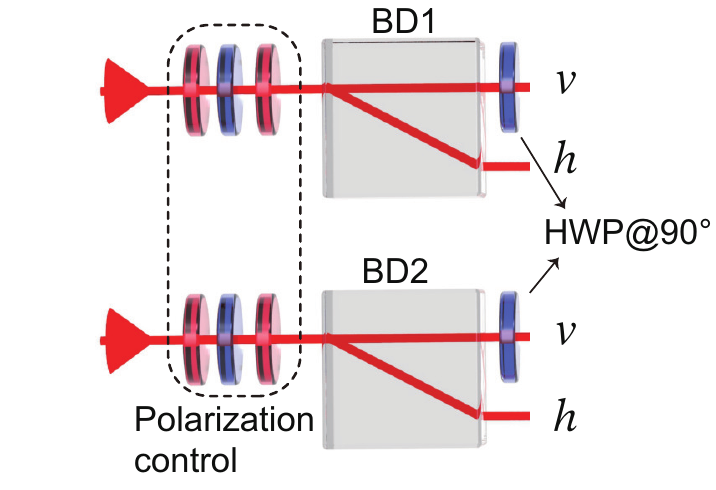} 
\caption{The left setup aims to generate photon pairs maximally entangled in the polarization degree of freedom (DOF) whereas the right setup aims to generate four-qubit GHZ state encoded in the polarization and path DOF.}
\label{Fig:PhotonSource}
\end{figure}
The pump beam is focused into the PPKTP crystal by two lenses with focal length of $\rm{f=75nm}$ and $\rm{f=125nm}$ (illustrated as $L_1$ in left of Fig.~\ref{Fig:PhotonSource}) with beam waist of $\rm{77\mu m}$. A dual-wave PBS splits the pump beam, which clockwise and counterclockwise pump the PPKTP crystal simultaneously. The PPKTP is placed into a homemade oven that is maintained at $\rm{29^\circ C}$ to achieve type-$\text{\RNum{2}}$ phase-matching condition of generating degenerated photons with wavelength at 810~nm. A dual-wave HWP is set in the counterclockwise path which transforms $\ket{V_p}\rightarrow\ket{H_p}$. The generated photons are superposed at PBS to create the maximally entangled photon pair in the form of $\ket{\Psi^+}=\frac{1}{\sqrt{2}}( \ket{HV}+ \ket{VH})$. The HWP at $45^\circ$ set before lens L3 transforms $\ket{\Psi^+}\rightarrow\ket{\Phi^+}=\frac{1}{\sqrt{2}}\left( \ket{HH}+\
\ket{VV}\right)$. The entangled photons are filtered by narrow-band filters(NBFs) and collected into single-mode fibres.\\

\subsection{Preparation of GHZ state}
The polarization-entangled photons are sent into two beam displacers (BDs) to generate $\ket{\text{GHZ}_4}$. We set a HWP sandwiched by two QWPs to correct the unitary transformations caused by fibres as shown in the right of Fig.~\ref{Fig:PhotonSource}. The BD is with the size of $\rm{10\times10\times28.3 mm^3}$, and can separate two polarization by 3~mm. The process to generate target state from ideal $\ket{\Phi^+}$ is 
\begin{equation}
\begin{split}
  \ket{\Phi^+}&=\frac{1}{\sqrt{2}}(\ket{VV}+\ket{HH}\\
  &\xrightarrow{\rm{BD1}}\frac{1}{\sqrt{2}}(\ket{VvV}+\ket{HhH})\\
 &\xrightarrow{\rm{BD2}}\frac{1}{\sqrt{2}}(\ket{VvVv}+\ket{HhHh}).
 \end{split}
\end{equation}
 
\begin{figure}[h]
    \centering
    \includegraphics[width = 0.95\textwidth]{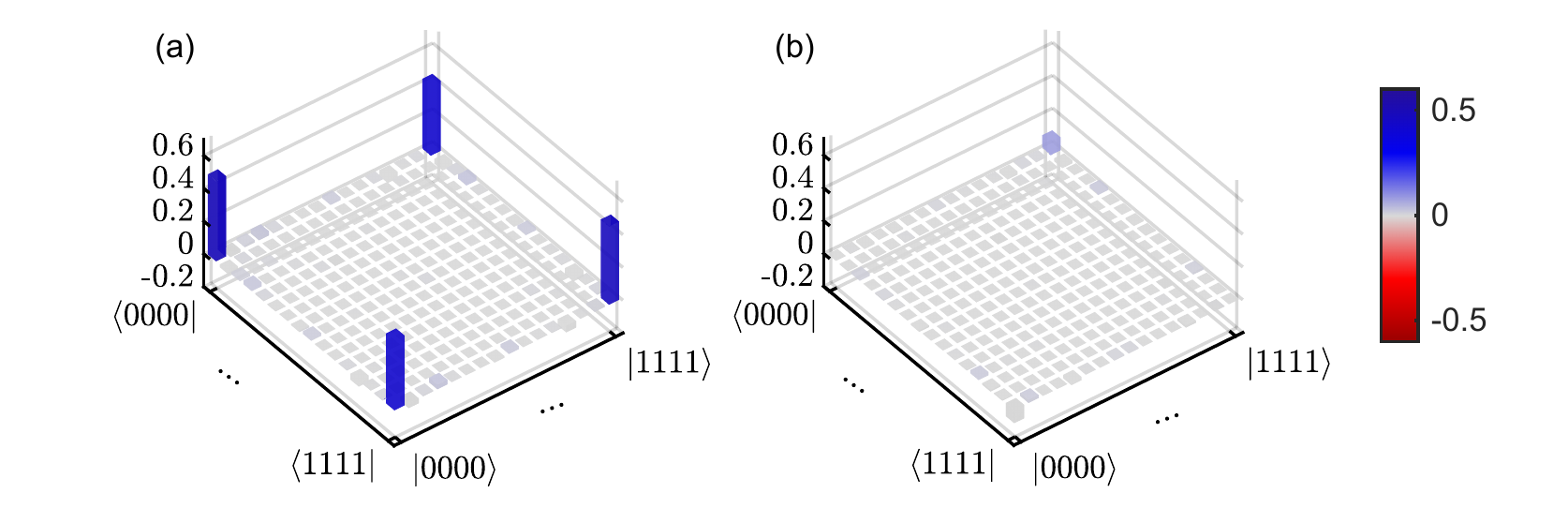}
    \caption{The density matrix of prepared $\rho_{\text{exp}}^{\text{GHZ}_4}$ is reconstituted by QST with $7.67424\times10^5$ coincidences. (a) is the real part of $\rho_{\text{exp}}^{\text{GHZ}_4}$, and (b) is the imaginary part of $\rho_{\text{exp}}^{\text{GHZ}_4}$.}
    \label{fig:TOMO_rho}
\end{figure}

We reconstructed the prepared $\rho_{\text{exp}}^{\text{GHZ}_4}$ using quantum state tomography (QST) technology using $7.67424\times 10^5$ coincidences, and its density matrix is shown in Fig.~\ref{fig:TOMO_rho}. We calculate the state fidelity by $\mathcal F=\tr(\sqrt{\sqrt{\rho_{\text{exp}}^{\text{GHZ}_4}}\rho_{\text{GHZ}_4} \sqrt{\rho_{\text{exp}}^{\text{GHZ}_4}}})$, and obtain the state fidelity of $\mathcal F=0.9546\pm 0.0006$. The main imperfections come from the high-order emission in the spontaneous down conversion process in PPKTP crystal and the interference on BD. We observe that the visibility of polarization-entangled photon pair is 0.98 (0.97) in $\ket{H}/\ket{V}$ ($\ket{+}/\ket{-}$) basis, and the visibility of interference on BD is 0.98. Other optical elements, such as HWP, QWP and PBS, are with high operation fidelity over 99\%.

\subsection{Implementation and detection of classical shadow}
\label{appendix:CS_exp}
\begin{figure}[h]
    \centering
    \includegraphics[width=0.35\columnwidth]{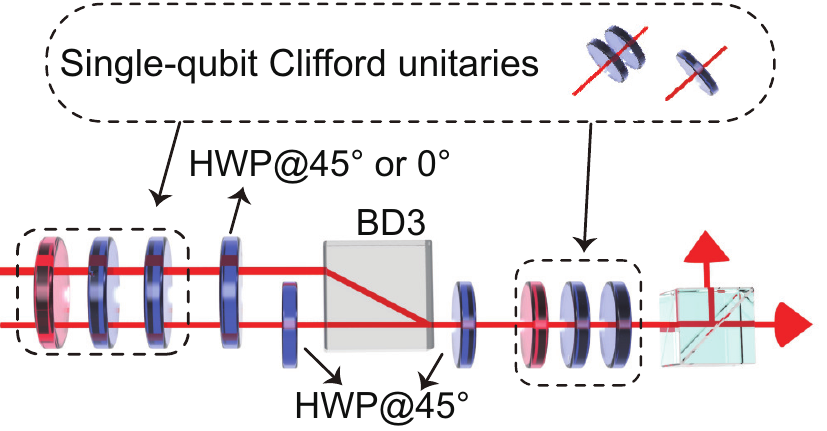}\hspace{1cm}
    \includegraphics[width=0.35\columnwidth]{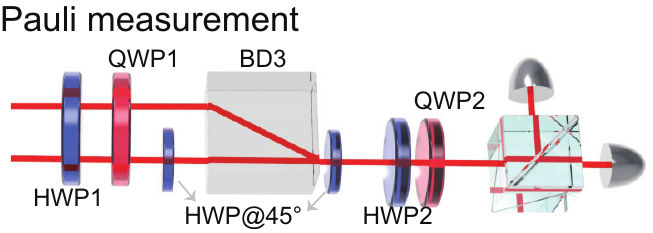} 
    \caption{The left setup aims to apply the single-qubit Clifford unitary and perform measurement in Pauli Z basis. The right setup aims to perform measurement in Pauli basis.}
    \label{fig:Measurement}
\end{figure}

The classical shadow in~\ref{eq:shadow_k} requires the single-qubit Clifford unitary $U_i$ acting on $i$th qubit and its corresponding outcome $\ket{\bmat_i}$ from projective measurements on Pauli Z basis. The single-qubit Clifford unitaries are realized on polarization-encoded qubit and the path-encoded qubit separately, which are written as
\begin{align}\label{eq:Clifford}
    U_{\text{POL}}=\left( \begin{matrix}
	U_{11}&		U_{12}\\
	U_{21}&		U_{22}\\
\end{matrix} \right)
,    U_{\text{PAT}}=\left( \begin{matrix}
	u_{11}&		u_{12}\\
	u_{21}&		u_{22}\\
\end{matrix} \right)
\end{align}
and satisfy $U_{\text{POL}}U_{\text{POL}}^{\dag}=I,\,\, U_{\text{PAT}}U_{\text{PAT}}^{\dag}=I$. Without loss of generality, the arbitrary single-photon hybrid state is $(\alpha\ket{H}+\beta\ket{V})\otimes(\gamma\ket{h}+\delta\ket{v}$. By applying the single-qubit Clifford unitaries, the output state is 
\begin{equation}\label{Eq:hybridstate}
    U_{\text{POL}}\begin{pmatrix}
    \alpha \\\beta
    \end{pmatrix}_{\text{POL}}\otimes
     U_{\text{PAT}}\begin{pmatrix}
    \gamma \\\delta
    \end{pmatrix}_{\text{PAT}}
    =
    \begin{pmatrix}
    (U_{11}\alpha+U_{12}\beta)(u_{11}\gamma+u_{12}\delta)\\
    (U_{11}\alpha+U_{12}\beta)(u_{21}\gamma+u_{22}\delta)\\
    (U_{21}\alpha+U_{22}\beta)(u_{11}\gamma+u_{12}\delta)\\
    (U_{21}\alpha+U_{22}\beta)(u_{21}\gamma+u_{22}\delta)\\
    \end{pmatrix}
\end{equation}
The measurement in Pauli Z basis yields $\ket{Hh}$, $\ket{Hv}$, $\ket{Vh}$ and $\ket{Vv}$ with corresponding probabilities $\abs{ (U_{11}\alpha+U_{12}\beta)(u_{11}\gamma+u_{12}\delta)}^2$, $\abs{(U_{11}\alpha+U_{12}\beta)(u_{21}\gamma+u_{22}\delta)}^2$, $\abs{(U_{21}\alpha+U_{22}\beta)(u_{11}\gamma+u_{12}\delta)}^2$ and $\abs{(U_{21}\alpha+U_{22}\beta)(u_{21}\gamma+u_{22}\delta)}^2$. Our experimental setup for implementing such single-qubit Clifford unitary and Pauli Z measurement is shown in the left of~\ref{fig:Measurement}, while the associated theoretical calculations are shown in~\ref{Eq:csexp}. 
\begin{equation}\label{Eq:csexp}
\begin{split}
  &(\alpha\ket{H}+\beta\ket{V})\otimes(\gamma\ket{h}+\delta\ket{v})\\
  &\xrightarrow{U_{\text{POL}}}[(U_{11}\alpha+U_{12}\beta)\ket{H}+(U_{21}\alpha+U_{22}\beta)\ket{V}]\otimes(\gamma\ket{h}+\delta\ket{v}),\\
 &\xrightarrow[\text{on both paths}]{\rm{HWP}}
 \left\{  
 \begin{array}{rcl}  
0^\circ,   &[(U_{11}\alpha+U_{12}\beta)\ket{H}-(U_{21}\alpha+U_{22}\beta)\ket{V}]\otimes(\gamma\ket{h}+\delta\ket{v})  \\
45^\circ,  &[(U_{11}\alpha+U_{12}\beta)\ket{V}+(U_{21}\alpha+U_{22}\beta)\ket{H}]\otimes(\gamma\ket{h}+\delta\ket{v})
\end{array}  
\right. \\
&\xrightarrow[\text{on path}\,\,h]{\rm{HWP@45^\circ}}
\left\{  
\begin{array}{rcl}  
0^\circ,   &\gamma[(U_{11}\alpha+U_{12}\beta)\ket{V}-(U_{21}\alpha+U_{22}\beta)\ket{H}]\ket{h}+\delta[(U_{11}\alpha+U_{12}\beta)\ket{H}-(U_{21}\alpha+U_{22}\beta)\ket{V}]\ket{v}  \\
45^\circ,  &\gamma[(U_{11}\alpha+U_{12}\beta)\ket{H}+(U_{21}\alpha+U_{22}\beta)\ket{V}]\ket{h}+\delta[(U_{11}\alpha+U_{12}\beta)\ket{V}+(U_{21}\alpha+U_{22}\beta)\ket{H}]\ket{v}
\end{array}  
\right. \\
&\xrightarrow[\text{postselect modes}\,\,Vh\,\,\text{and}\,\,Hv]{\rm{BD3}}
\left\{  
\begin{array}{rcl}  
0^\circ,   &\gamma(U_{11}\alpha+U_{12}\beta)\ket{V}+\delta(U_{11}\alpha+U_{12}\beta)\ket{H}  \\
45^\circ,  &\gamma(U_{21}\alpha+U_{22}\beta)\ket{V}+\delta(U_{21}\alpha+U_{22}\beta)\ket{H}
\end{array}  
\right. \text{(We omit the path information)}\\
&\xrightarrow{\rm{HWP@45^\circ}}
\left\{  
\begin{array}{rcl}  
0^\circ,    &\gamma(U_{11}\alpha+U_{12}\beta)\ket{H}+\delta(U_{11}\alpha+U_{12}\beta)\ket{V} \\
45^\circ,   &\gamma(U_{21}\alpha+U_{22}\beta)\ket{H}+\delta(U_{21}\alpha+U_{22}\beta)\ket{V}
\end{array}  
\right. \\
&\xrightarrow{U_{\text{PAT}}}
\left\{  
\begin{array}{rcl}  
0^\circ,    &(U_{11}\alpha+U_{12}\beta)(u_{11}\gamma+u_{12}\delta)\ket{H}+(U_{11}\alpha+U_{12}\beta)(u_{21}\gamma+u_{22}\delta)\ket{V} \\
45^\circ,   &(U_{21}\alpha+U_{22}\beta)(u_{11}\gamma+u_{12}\delta)\ket{H}+(U_{21}\alpha+U_{22}\beta)(u_{21}\gamma+u_{22}\delta)\ket{V}
\end{array}  
\right. \\
&\xrightarrow{\rm{PBS}}
\left\{  
\begin{array}{rcl}  
&0^\circ, \text{transmitted},  &\text{results of}\,\,\ket{Hh}\,\,\text{with probability of\,\,} \abs{(U_{11}\alpha+U_{12}\beta)(u_{11}\gamma+u_{12}\delta)}^2 \\
&0^\circ, \text{reflected},    &\text{results of}\,\,\ket{Hv}\,\,\text{with probability of\,\,}\abs{(U_{11}\alpha+U_{12}\beta)(u_{21}\gamma+u_{22}\delta)}^2 \\
&45^\circ,\text{transmitted},  &\text{results of}\,\,\ket{Vh}\,\,\text{with probability of\,\,}\abs{(U_{21}\alpha+U_{22}\beta)(u_{11}\gamma+u_{12}\delta)}^2 \\
&45^\circ,\text{reflected},     &\text{results of}\,\,\ket{Vv}\,\,\text{with probability of\,\,}\abs{(U_{21}\alpha+U_{22}\beta)(u_{21}\gamma+u_{22}\delta)}^2
\end{array}  
\right. \\
\end{split}
\end{equation}
Thus, the probabilities of observing photon after PBS is $\abs{ (U_{11}\alpha+U_{12}\beta)(u_{11}\gamma+u_{12}\delta)}^2$, $\abs{(U_{11}\alpha+U_{12}\beta)(u_{21}\gamma+u_{22}\delta)}^2$, $\abs{(U_{21}\alpha+U_{22}\beta)(u_{11}\gamma+u_{12}\delta)}^2$ and $\abs{(U_{21}\alpha+U_{22}\beta)(u_{21}\gamma+u_{22}\delta)}^2$, which is the same as~\ref{Eq:hybridstate}. 

The set of single-qubit Clifford unitary is shown in~\ref{tab:Clifford unitaries}, associated with its angle setting of waveplates. Note that any single-qubit Clifford unitary can be realized with up to three waveplates. Some specific unitary can be realized with one HWP.   

\begin{table} 
    \centering
    \begin{tabular}{|c|c c c|} 
    \hline
    Single-qubit Clifford unitaries & QWP$_1$ & HWP$_2$ & HWP$_3$\\
    \hline\hline
    Pauli oprations & \,\, & \,\, & \,\,\\
    \hline
    $I$ & \,\, & $0^\circ$ & $0^\circ$\\
    $X$ & \,\, & \,\, & $45^\circ$\\
    $Y$ & \,\, &  $0^\circ$  & $45^\circ$\\
    $Z$ & \,\, & \,\,& $0^\circ$\\
    \hline
    $2\pi/3$ rotations & \,\, & \,\, & \,\,\\
    \hline
    $e^{(\frac{-i\pi}{2}) ( \frac{X}{2})}\cdot e^{( \frac{-i\pi}{2})( \frac{Y}{2})}
$&$90^\circ$ & $0^\circ$& $22.5^\circ$\\
 $e^{(\frac{-i\pi}{2}) ( \frac{X}{2})}\cdot e^{( \frac{-i\pi}{2})( \frac{-Y}{2})}
$&$0^\circ$ & $22.5^\circ$& $0^\circ$\\
 $e^{(\frac{-i\pi}{2}) ( \frac{-X}{2})}\cdot e^{( \frac{-i\pi}{2})( \frac{Y}{2})}
$&$0^\circ$ & $0^\circ$& $22.5^\circ$\\
 $e^{(\frac{-i\pi}{2}) ( \frac{-X}{2})}\cdot e^{( \frac{-i\pi}{2})( \frac{-Y}{2})}
$&$90^\circ$ & $22.5^\circ$& $0^\circ$\\
 $e^{(\frac{-i\pi}{2}) (\frac{Y}{2})}\cdot e^{( \frac{-i\pi}{2})( \frac{X}{2})}
$&$135^\circ$ & $0^\circ$& $22.5^\circ$\\
 $e^{(\frac{-i\pi}{2}) (\frac{Y}{2})}\cdot e^{( \frac{-i\pi}{2})( \frac{-X}{2})}
$&$45^\circ$ & $0^\circ$& $22.5^\circ$\\
$e^{(\frac{-i\pi}{2}) (\frac{-Y}{2})}\cdot e^{( \frac{-i\pi}{2})( \frac{X}{2})}
$&$135^\circ$ & $22.5^\circ$& $0^\circ$\\
$e^{(\frac{-i\pi}{2}) (\frac{-Y}{2})}\cdot e^{( \frac{-i\pi}{2})( \frac{-X}{2})}
$&$45^\circ$ & $22.5^\circ$& $0^\circ$\\
    \hline
    $\pi/2$ rotations& \,\, & \,\, & \,\,\\
    \hline
$e^{(\frac{-i\pi}{2}) (\frac{X}{2})}$&$135^\circ$ & $0^\circ$& $0^\circ$\\
$e^{(\frac{-i\pi}{2}) (\frac{-X}{2})}$
&$45^\circ$ & $0^\circ$& $0^\circ$\\
$e^{(\frac{-i\pi}{2}) (\frac{Y}{2})}$
&\,\, & $0^\circ$& $22.5^\circ$\\
$e^{(\frac{-i\pi}{2}) (\frac{-Y}{2})}$&\,\, & $0^\circ$& $67.5^\circ$\\
$e^{(\frac{-i\pi}{2}) (\frac{-X}{2})}\cdot e^{(\frac{-i\pi}{2})(\frac{Y}{2})}\cdot e^{(\frac{-i\pi}{2}) (\frac{X}{2})}$&$0^\circ$ & $0^\circ$& $0^\circ$\\
$e^{(\frac{-i\pi}{2}) (\frac{-X}{2})}\cdot e^{(\frac{-i\pi}{2}) (\frac{-Y}{2})}\cdot
e^{(\frac{-i\pi}{2}) (\frac{X}{2})}$&$90^\circ$ & $0^\circ$& $0^\circ$\\
     \hline
    Hadamard-like & \,\, & \,\, & \,\,\\
    \hline
$e^{(\frac{-i\pi X}{2})}\cdot e^{( \frac{-i\pi}{2})( \frac{Y}{2})}$&\,\,& \,\,& $22.5^\circ$\\
$e^{(\frac{-i\pi X}{2})}\cdot e^{( \frac{-i\pi}{2})( \frac{-Y}{2})}$
&\,\,& \,\,& $67.5^\circ$\\
$e^{(\frac{-i\pi Y}{2})}\cdot e^{( \frac{-i\pi}{2})( \frac{X}{2})}$
&$135^\circ$&$0^\circ$& $45^\circ$\\
$e^{(\frac{-i\pi Y}{2})}\cdot e^{( \frac{-i\pi}{2})( \frac{-X}{2})}$
&$45^\circ$&$0^\circ$& $45^\circ$\\
$e^{(\frac{-i\pi}{2}) ( \frac{X}{2})}\cdot e^{( \frac{-i\pi}{2})( \frac{Y}{2})}\cdot e^{( \frac{-i\pi}{2})( \frac{X}{2})}$&$90^\circ$&$0^\circ$& $45^\circ$\\
$e^{(\frac{-i\pi}{2}) ( \frac{-X}{2})}\cdot e^{( \frac{-i\pi}{2})(\frac{Y}{2})}\cdot e^{( \frac{-i\pi}{2})( \frac{-X}{2})}$
&$0^\circ$&$0^\circ$& $45^\circ$\\
    \hline
    \end{tabular}
    \caption{The set of single-qubit Clifford unitary~\cite{Webb2016} and its experimental setting. Here the subscripts indicate the order of the wave-plates in the left of~\ref{fig:Measurement}.
    {There are other forms of waveplates combination and choice of angles for implementing the Clifford unitaries. For the sake of simplicity in our experimental operations, we have chosen this form in the table}.}
    \label{tab:Clifford unitaries}
\end{table}
 
\subsection{Measurement in Pauli basis}
 The setup to perform measurement in Pauli basis is shown in the right of~\ref{fig:Measurement}. Similarly, the input single-photon hybrid state can be arbitrary, i.e., $(\alpha\ket{H}+\beta \ket{V})\otimes (\gamma\ket{h}+\delta\ket{v})$. The measurement on polarization-encoded qubit is realised by $\rm{HWP_1@\theta_1}$ and $\rm{QWP_1@\vartheta_1}$, while the measurement on path-encoded qubit is realised  by the $\rm{HWP_2@\theta_2}$ and  $\rm{QWP_2@\vartheta_2}$.
By choosing the appropriate angle for the $\rm{HWPs}$ and $\rm{QWPs}$, we can perform measurement on arbitrary basis including the Pauli basis. The process is described as 
\begin{equation}
\begin{split}
&(\alpha\ket{H}+\beta \ket{V})\otimes(\gamma\ket{h}+\delta\ket{v})\\
&\xrightarrow[\rm{QWP_1@\vartheta_1}]{\rm{HWP_1@\theta_1}}\ket{H}\otimes(\gamma\ket{h}+\delta\ket{v})\\
&\xrightarrow[\rm{on\,\,path\,\,}h]{\rm{HWP@45^\circ}}\gamma\ket{Vh}+\delta\ket{Hv}\\
&\xrightarrow[\text{postselect modes}\,\,Vh\,\,\text{and}\,\,Hv]{\rm{BD3}}\gamma\ket{V}+\delta\ket{H}\text{(We omit the path information)}\\
&\xrightarrow{\rm{HWP@45^\circ}}\gamma\ket{H}+\delta\ket{V}\\
&\xrightarrow[\rm{QWP_2@\vartheta_2}]{\rm{HWP_2@\theta_2}}\ket{H}\\ 
&\xrightarrow{\rm{PBS}}
\left\{  
\begin{array}{rcl}  
\text{transmitted}, &\text{results of}\,\,(\alpha\ket{H}+\beta \ket{V})\otimes(\gamma\ket{h}+\delta\ket{v}) \\
\text{reflected},   &\text{resutls of}\,\,(\alpha\ket{H}+\beta \ket{V})\otimes(\gamma\ket{h}+\delta\ket{v})^\perp 
\end{array}  
\right., 
\end{split}
\end{equation}
where $(\gamma\ket{h}+\delta\ket{v})^\perp$ is the orthogonal state of $\gamma\ket{h}+\delta\ket{v}$.

\subsection{Data acquisition}
\label{appendix:Data}
\begin{figure}[htp]
\centering
\includegraphics[scale=1]{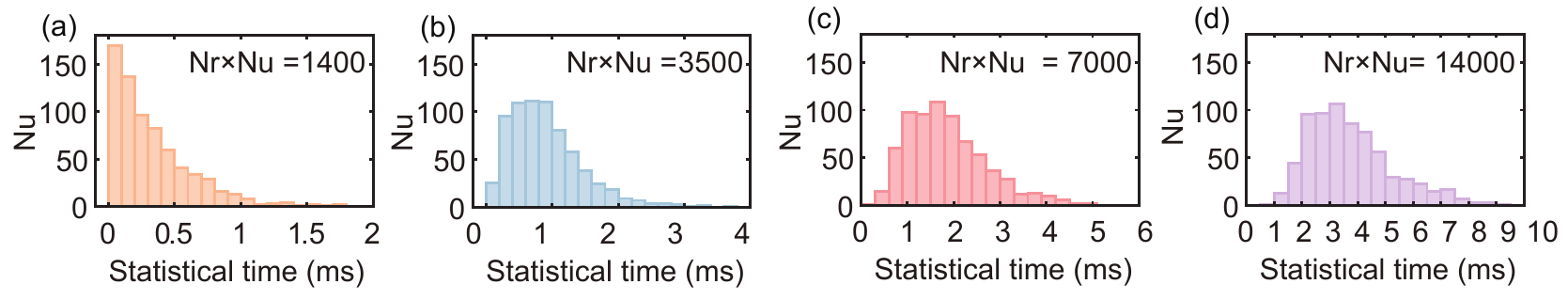}
\caption{The histogram of statistical time for accumulating (a) $N_r=2$, (b) $N_r=5$, (c) $N_r=10$ and (d) $N_r=20$ coincidences for each unitary, respectively. The y-axis $N_u$ represents the number of unitarites. The all coincidence is calculated by $700N_r$.}
\label{Fig:measurement time}
\end{figure}

One key ingredient of CS algorithm is to construct classical shadow in~\ref{eq:shadow_k}. The single-qubit Clifford unitary and measurement in Pauli Z basis are discussed in~\ref{appendix:CS_exp}. In this section, we discuss the method to extract $\ket{\bmat_i}$ from the collected data. The photons are recorded by time-correlated single-photon counting (TCSPC) system, which tags the arriving time of each photon. We fix the time tag of one photon, then search the time tag of the other photon that fall in a 3~ns time window (coincidence window). If there does not exist such a coincidence, we skip to the next time tag and repeat the process above until we obtain one coincidence. Experimentally, we randomly select unitaries $U$ from~\ref{tab:Clifford unitaries}, each of which we extract $N_r$ coincidences. The histogram of statistical time for accumulating $N_r$ coincidences of 700 unitaries is shown in~\ref{Fig:measurement time}. 

\begin{table}
\centering
\begin{tabular}{|c|c|c|c|c| }
    \hline
    \multicolumn{5}{c}{Observables}\\

    \hline\hline
$Y\otimes I\otimes I\otimes Z$&$I\otimes Y\otimes Z\otimes I$&$I\otimes Z\otimes I\otimes I$&$I\otimes Z\otimes I\otimes Z$&$X\otimes I\otimes I\otimes X$\\
$I\otimes Z\otimes Y\otimes I$&$Z\otimes X\otimes I\otimes I$&$I\otimes I\otimes Z\otimes X$&$I\otimes I\otimes X\otimes Z$&$X\otimes I\otimes I\otimes Z$\\
$Z\otimes Y\otimes I\otimes I$&$X\otimes I\otimes I\otimes X$&$X\otimes I\otimes I\otimes I$&$I\otimes Y\otimes I\otimes I$&$I\otimes Y\otimes I\otimes Z$\\
$I\otimes X\otimes X\otimes I$&$I\otimes X\otimes I\otimes Z$&$I\otimes I\otimes Y\otimes Z$&$I\otimes Y\otimes Z\otimes I$&$I\otimes I\otimes X\otimes I$\\
$Y\otimes I\otimes I\otimes Z$&$Z\otimes I\otimes I\otimes I$&$Y\otimes Z\otimes I\otimes I$&$I\otimes Y\otimes Y\otimes I$&$I\otimes I\otimes I\otimes Z$\\
$Z\otimes I\otimes I\otimes X$&$I\otimes Z\otimes I\otimes Y$&$I\otimes I\otimes Y\otimes Z$&$X\otimes I\otimes I\otimes I$&$I\otimes Y\otimes Y\otimes I$\\
$I\otimes I\otimes Y\otimes I$&$Y\otimes I\otimes X\otimes I$&$Y\otimes I\otimes X\otimes I$&$X\otimes I\otimes I\otimes I$&$I\otimes I\otimes X\otimes X$\\
$I\otimes X\otimes I\otimes Z$&$I\otimes Y\otimes I\otimes I$&$X\otimes I\otimes I\otimes Z$&$I\otimes I\otimes Z\otimes I$&$X\otimes I\otimes I\otimes I$\\
$I\otimes X\otimes I\otimes X$&$I\otimes Y\otimes Y\otimes I$&$I\otimes X\otimes Y\otimes I$&$X\otimes I\otimes X\otimes I$&$Y\otimes X\otimes I\otimes I$\\
$I\otimes I\otimes X\otimes I$&$Z\otimes I\otimes I\otimes I$&$I\otimes Y\otimes Z\otimes I$&$I\otimes I\otimes I\otimes X$&$I\otimes I\otimes I\otimes X$\\
\hline
\end{tabular}
 \caption{The 50 local observables $\Omat_l$ that are experimentally estimated.}
 \label{tab:50observable}
\end{table}

\section{Experimental results}
\subsection{Statistical error of the estimation}
\begin{figure}[htb]
    \centering
    \includegraphics[width = 0.95\textwidth]{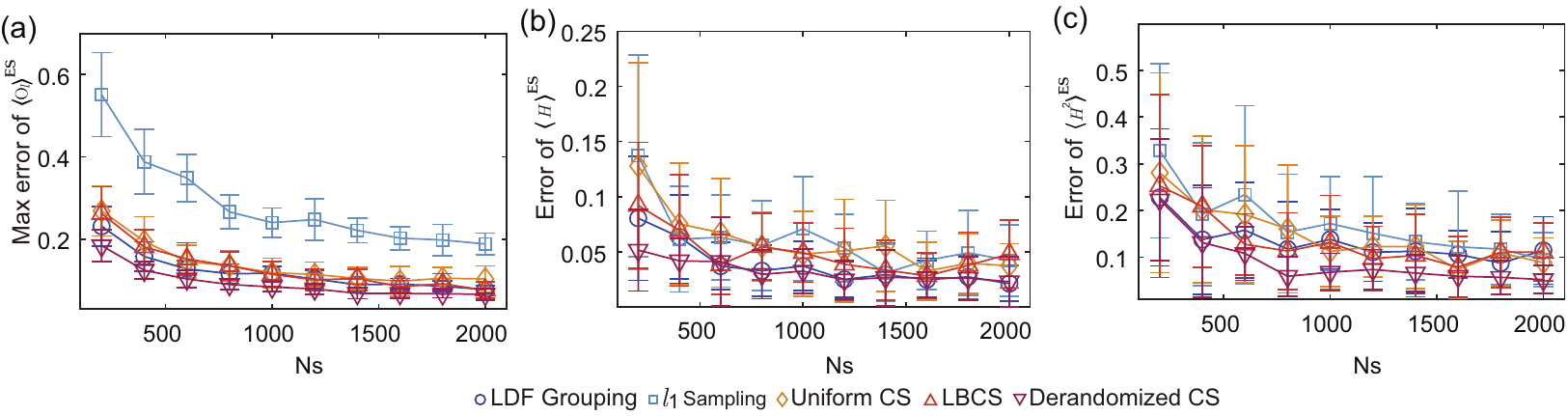}
    \caption{The results of estimation of observables with statistical errors. (a), (b) and (c) is the results with statistical errors that correspond to Fig.~2(a), Fig.~2(c) and Fig.~2(d) in the main text, respectively. The errorbar is the standard deviation of estimation error over 20 independent repetitions of the entire setup, in which we fixed $N_r = 5$ in each measurement basis.}
    \label{fig:Errorbar_observable}
\end{figure}

In the main text, we show the maximum errors for the estimation of the $50$ local observables that are tensor products of Pauli operators acting non-trivially on maximally two qubits. The local observables are exhibited in~\ref{tab:50observable}. 
The statistical errors of the results in Fig. 2 in main text are shown in~\ref{fig:Errorbar_observable}.
We calculate the standard deviation of the estimated $\ES{\Olmat}$, $\ES{{H}}$, and $\ES{{H^2}}$ over $20$ independent repetitions of the entire setup. In what follows, we use the same parameter set-up for different tasks to keep consistency if not clarified.
Note that with each measurement basis, we could increase the number of samples by accumulating $N_r>1$ coincidences to improve the statistics. 
In Fig. 2(b) in the main text,  we display the maximum absolute error of the observables in an ascending order to show the error dependence of the number of observables.
\subsection{Numerical simulation with noiseless state}
\begin{figure}[h!t]
    \centering
    \includegraphics[width =0.95\textwidth]{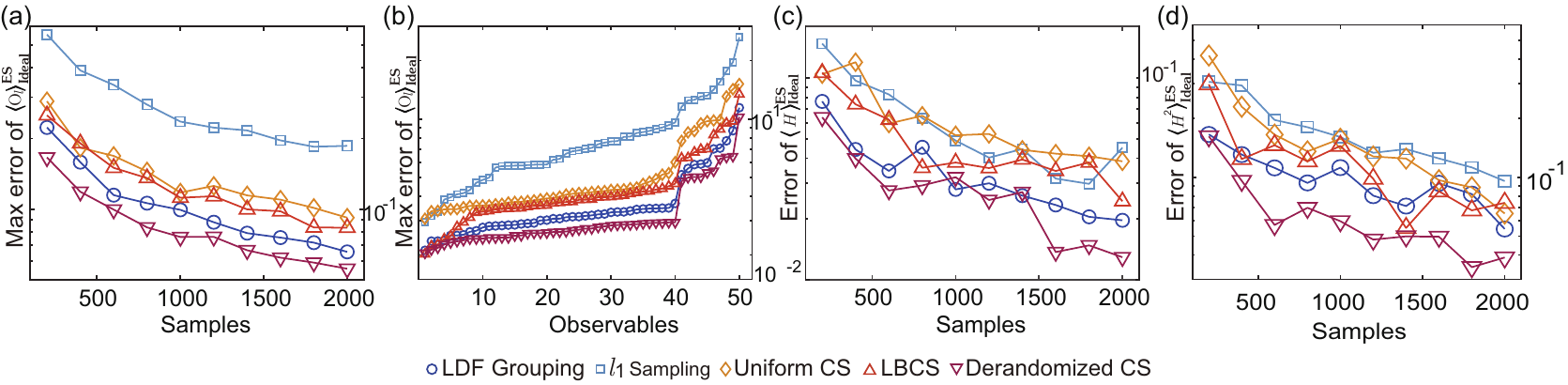}
    \caption{The results for numerical simulation with noiseless state $\ket{\text{GHZ}_4}$. (a) The maximum absolute error of $\ES{\Olmat}_{\text{ideal}}$ over $50$ local observables $\Olmat$ that are randomly selected from the Pauli set with different number of samples $N_s$.
	(b) The maximum absolute error of $\ES{\Olmat}_{\text{ideal}}$ with different number of local observables, each of which we fix $N_s=2000$, and we collect $N_r = 5$ coincidences for each sample.
	(c) and (d) are the errors of estimated energy $\ES{H}_{\text{ideal}}$ and that of estimated Hamiltonian moment $\ES{H^2}_{\text{ideal}}$ with different $N_s$, respectively.
	}
    \label{fig:Observable_ideal}
\end{figure}

In the main text, we observe that the derandomized CS method outperforms other schemes in the estimation of observables, especially in the estimation of $\braket{H^2}$. The main reason is that the second moment of the Hamiltonian $H^2$ has more (non-commuting) terms than $H$, which makes $H^2$ more costly to measure. This conclusion can also be reflected by the the numerical simulation with noiseless state $\ket{\text{GHZ}_4}$. We denote the estimation with ideal state as $\ES{\bullet}_{\text{ideal}}$, and the results for error of estimations are shown in~\ref{fig:Observable_ideal}. The error here is defined as the difference between $\ES{\bullet}_{\text{ideal}}$ and the direct calculation with $\ket{\text{GHZ}_4}$. As shown in~\ref{fig:Observable_ideal}, we observe that derandomized CS method outperforms other schemes as well. We  expect that the advanced measurement scheme could show more advantages when the system size increases or the Hamiltonian becomes complex, as indicated in Refs.~\cite{huang2021efficient,wu2021overlapped,hadfield2021adaptive}. 

\subsection{Estimation of cluster Hamiltonian with Ising interactions}
\begin{figure}[tb]
    \centering
    \includegraphics[width = 0.65\textwidth]{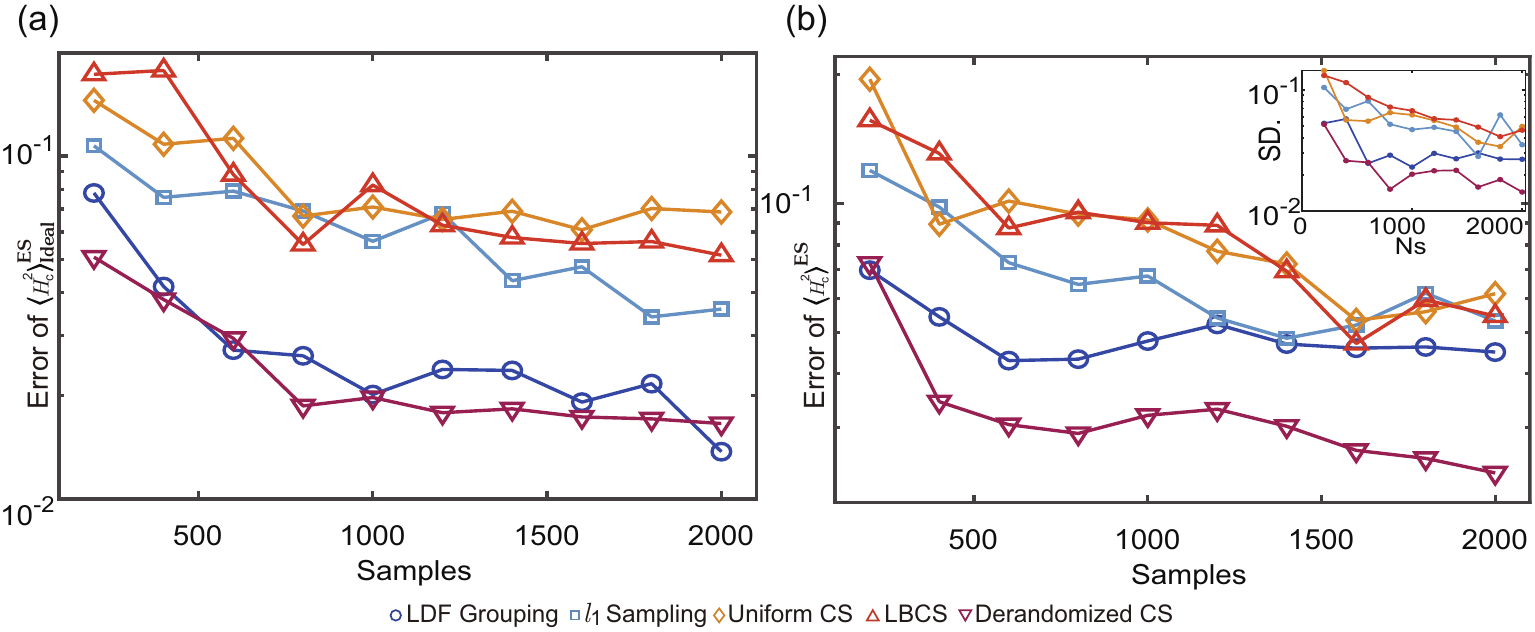}
        \caption{The error of estimated Hamiltonian moment of $H^2$. (a) The estimation of $\ES{H^2}_{\text{ideal}}$ with noiseless state $\ket{\text{GHZ}_4}$. (b) The estimation of $\ES{H^2}$ with experimentally prepared state $\rho_{\text{exp}}^{\text{GHZ}_4}$.  The inset shows the standard deviation (SD) of estimation error over 20 independent repetitions of the
entire setup.}
    \label{fig:C_Hamiltonian}
\end{figure}

We also consider a cluster Hamiltonian with Ising interactions in the form of $H = H_C + H_I$ with periodic boundary conditions. Here, $H_C = J\sum_j Z_jX_{j+1}Z_{j+2}$ is the cluster Hamiltonian, which has $\mathbb{Z}_2 \times \mathbb{Z}_2$ global symmetry, and $H_I = h_1 \sum_j X_j + h_2 \sum_j Y_jY_{j+1}$ is the Ising interaction. 
The ground state in the cluster phase has symmetry protected topological order and is shown to have a continuous quantum phase transition as a competition between the cluster and Ising terms~\cite{son2011quantum,skrovseth2009phase}. 
We set the normalized  strength as $J =
h_1 = h_2 = 1/4$.
The results for second-order moment $H^2$ are shown in~\ref{fig:C_Hamiltonian}(a) (numerical simulation with noiseless state) and~\ref{fig:C_Hamiltonian}(b) (experimental results with $\rho_{\text{exp}}^{\text{GHZ}_4}$), respectively. Again, we observe the superiority of results with LDF grouping and derandomized CS scheme compared with other schemes. 

It is worth to mention that the derandomization  has a relatively small variance compared to LDF grouping. The variance is evaluated using Eq.~\eqref{eq:derand} and Eq.~\eqref{eq:var_group}, respectively.
Remarkably, compared to the ideal case, the performance of derandomized CS scheme is slightly better than LDF grouping in the experiments, as reflected in~\ref{fig:C_Hamiltonian}(b). 
This may be attributed to the fact that some outcomes from certain measurement  basis may have non-negligible deviations from the exact expectation. Therefore, the derandomization, which uses more measurement bases,  could be more robust to measurement noise in this case.
A  definite answer may be an interesting direction.

\subsection{Estimation of hydrogen molecular Hamiltonian}
\begin{figure}[ht]
    \centering
    \includegraphics[width = 0.85\textwidth]{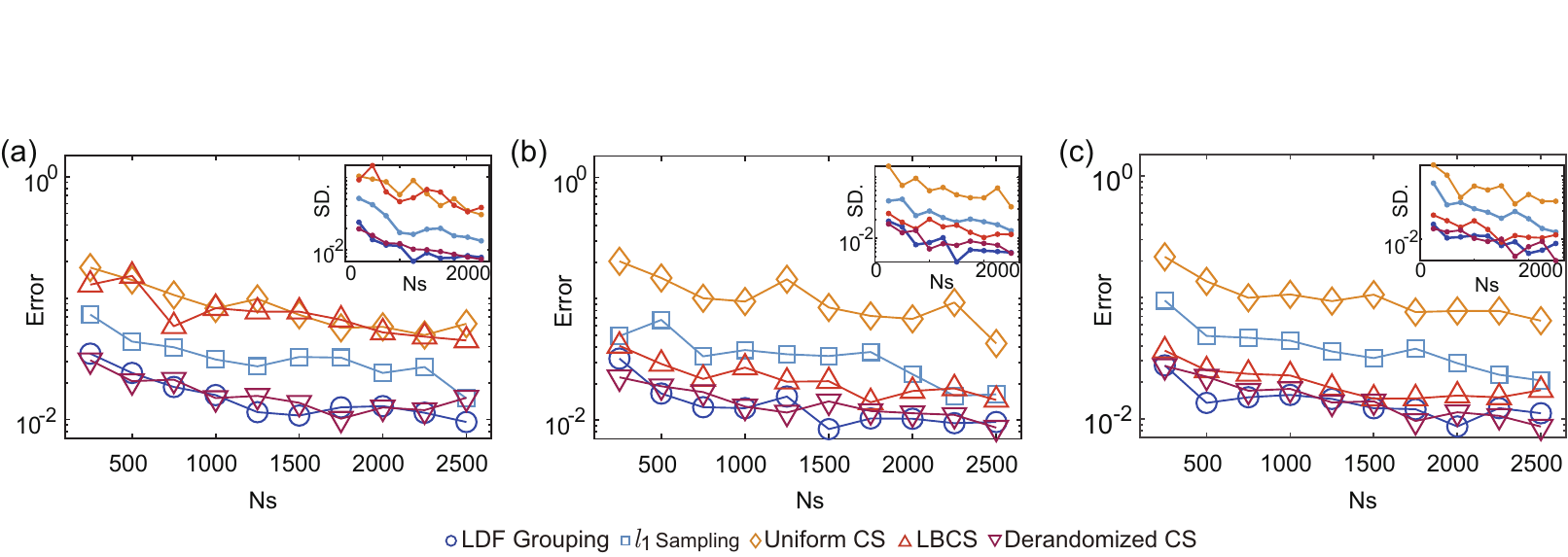}
    \caption{Energy estimation error of the hydrogen molecular. The Hamiltonian is represented in a minimal STO-3G basis with 4 spin orbitals, which is encoded in qubit ones under the fermion-to-qubit mappings: Jordan-Wigner (a), parity (b), and Bravyi-Kitaev (c). The inset shows the standard deviation for the estimation errors, which is calculated over 20 independent repetitions of the entire setup. The $N_r$ here is fixed as 5.}
    \label{fig:molecular}
\end{figure}

We next consider the energy estimation of hydrogen molecular.
Hydrogen molecular Hamiltonian is represented in a minimal STO-3G basis with 4 spin orbitals, which is encoded in qubits under the fermion-to-qubit mappings: Jordan-Wigner (JW), parity, and Bravyi-Kitaev (BK). 
We show the energy estimations from the experimentally prepared state $\rho_{\text{exp}}^{\text{GHZ}_4}$ with different encodings in Fig.~\ref{fig:molecular}.
As shown in Refs.~\cite{huang2021efficient,wu2021overlapped,hadfield2021adaptive}, based on the variance (except for derandomization) and estimation error analysis computed on the ground state of the four-qubit hydrogen molecular, five measurement schemes considered in the main text should have similar performance, aligning with the experimental results.
Nevertheless, one can expect that the advanced measurement schemes could significantly outperform the conventional measurements when  the problem size increases, as theoretically and numerically shown in the references~\cite{huang2021efficient,wu2021overlapped,hadfield2021adaptive}.

\begin{figure}[ht]
\centering
\includegraphics[width = 0.9\textwidth]{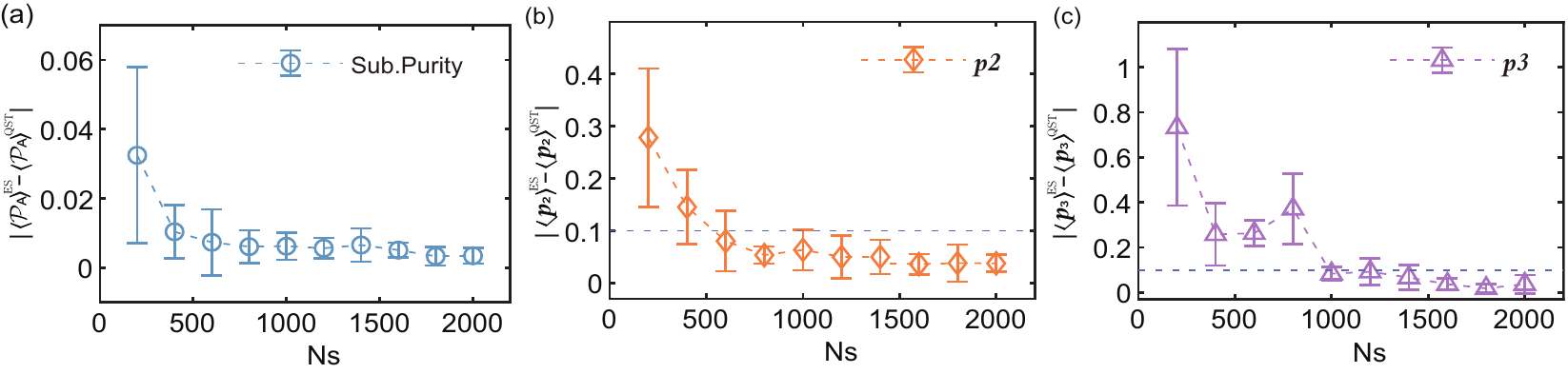}[htp]
\caption{ Estimation errors of (a) the subsystem purity $\mathcal P_A$,  (b) the $p_2$ moments, and (c) the $p_3$ moments with different number of samples. The standard deviation is given over 10 independent repetitions of the entire setup. }
\label{fig: Error_samples}
\end{figure}

\begin{figure}
    \centering
    \includegraphics[width = 0.9\textwidth]{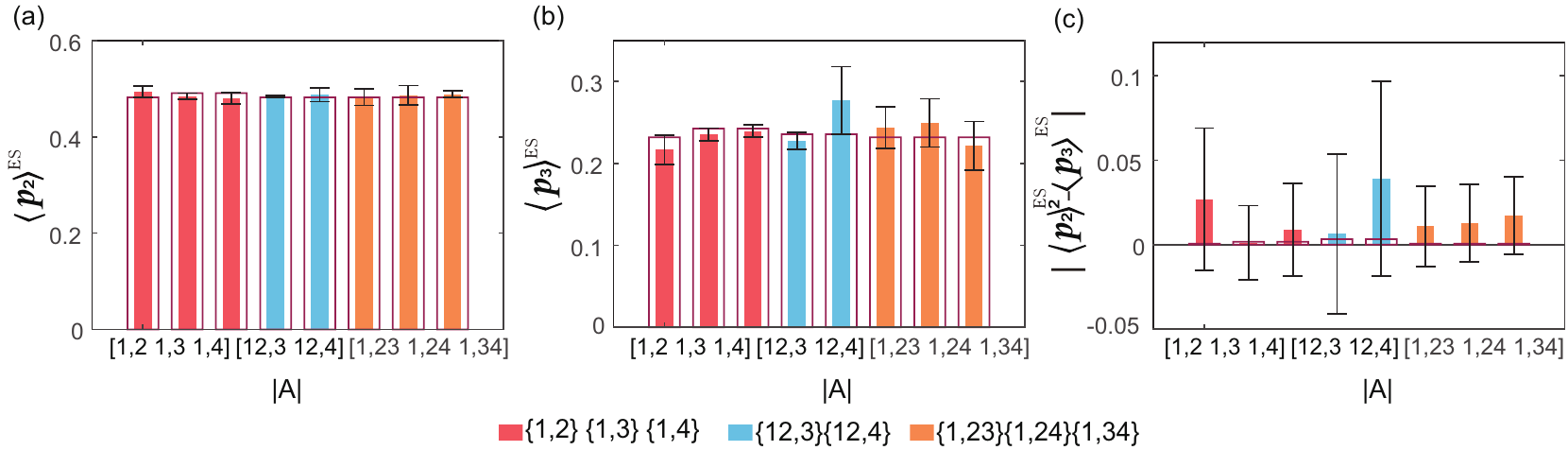}
    \caption{Estimation error of (a) the $p_2$ moments and (b) the $p_3$ moments in different  subsystem partitioning with the same samples $N_s=1000$. (c)  The estimation of ${p_2^2}-{p_3}$. The subsystem division is shown in the figure legend. The standard deviation is given over 5 independent repetitions of the entire setup.
    }
    \label{fig:PPT_mixed}
\end{figure}

\subsection{Estimation of nonlinear function}
Finally, we show the results of nonlinear function estimation considered in the main text. In Fig.~\ref{fig: Error_samples}, we show the estimation errors and the standard deviation of the subsystem purity $\mathcal P_A$,  $p_2$ and $p_3$ moments, with the subsystem division shown in the inset of Fig.~3 in the main text.
While in the main text we demonstrate the $p_3$-PPT condition for the full system, which is a pure state ideally,  one can use the $p_3$-PPT condition to detect the bipartite entanglement of a mixed state~\cite{elben2020mixed,neven2021symmetry}.
In Fig.~\ref{fig:PPT_mixed}, we illustrate the estimation of $p_2$ and $p_3$ for the reduced density matrix of the subsystem.  The subsystem division is displayed in the figure legend. 
Here, we show the estimation of PT-moments as a proof-of-principle demonstration; however, one cannot assure the violation of $p_3$-PPT condition, as shown in Fig.~\ref{fig:PPT_mixed}(c).
The entanglement structure could be inferred using the experimental results shown in the main text.

\end{document}